\documentclass[aps,prd,amsfonts,preprintnumbers,superscriptaddress,showpacs,nofootinbib]{revtex4}
\pdfoutput=1
\usepackage{graphics}
\usepackage{amsmath}
\usepackage{amssymb}
\usepackage{psfrag}
\usepackage{epsfig}
\usepackage{epsf}
\usepackage{float}
\usepackage{slashed}
\allowdisplaybreaks
\usepackage{dsfont}
\usepackage{amsfonts}
\usepackage{color}
\usepackage[
colorlinks=true,
linkcolor=black,
breaklinks=true,
urlcolor=blue,
citecolor=blue]{hyperref}

\newcommand{\be}{\begin{equation}}
\newcommand{\ee}{\end{equation}}
\newcommand{\ba}{\begin{eqnarray}}
\newcommand{\ea}{\end{eqnarray}}

\newcommand{\order}[1]{\mathcal{O}\left(#1\right)}

\numberwithin{equation}{section}
\renewcommand\theequation{\arabic{equation}}

\begin{document}

\title{{\boldmath$\theta$}-dependence of light nuclei and nucleosynthesis }

\author{Dean~Lee}
\email[]{Email: leed@frib.msu.edu}
\affiliation{Facility for Rare Isotope Beams and Department of Physics and
  Astronomy,  Michigan State
University, MI 48824, USA}

\author{Ulf-G.~Mei{\ss}ner}
\email[]{Email: meissner@hiskp.uni-bonn.de}
\affiliation{Helmholtz-Institut~f\"{u}r~Strahlen-~und~Kernphysik~and~Bethe~Center~for~Theoretical~Physics,
~Universit\"{a}t~Bonn,~D-53115~Bonn,~Germany}
\affiliation{Institute~for~Advanced~Simulation,~Institut~f\"{u}r~Kernphysik,
and J\"{u}lich~Center~for~Hadron~Physics, Forschungszentrum~J\"{u}lich,~D-52425~J\"{u}lich,~Germany}
\affiliation{Tbilisi State University, 0186 Tbilisi, Georgia}

\author{Keith~A.~Olive}
\email[]{Email: olive@umn.edu}
\affiliation{William I. Fine Theoretical Physics Institute,
School of Physics and Astronomy, University of Minnesota, Minneapolis, MN 55455, USA}

\author{Mikhail~Shifman}
\email[]{Email: shifman@umn.edu}
\affiliation{William I. Fine Theoretical Physics Institute, School of Physics and Astronomy, University of Minnesota, Minneapolis, MN 55455, USA}

\author{Thomas~Vonk}
\email[]{Email: vonk@hiskp.uni-bonn.de}
\affiliation{Helmholtz-Institut~f\"{u}r~Strahlen-~und~Kernphysik~and~Bethe~Center~for~Theoretical~Physics,
~Universit\"{a}t~Bonn,~D-53115~Bonn,~Germany}

\date{\today}

\begin{abstract}
  We investigate the impact of the QCD vacuum at nonzero $\theta$ on the properties of light nuclei, Big Bang
  nucleosynthesis, and stellar nucleosynthesis.  Our analysis starts with a calculation of the $\theta$-dependence
  of the neutron-proton mass difference and neutron decay using chiral perturbation theory.  We then discuss the
  $\theta$-dependence of the nucleon-nucleon interaction using a one-boson-exchange model and compute the
  properties of the two-nucleon system.  Using the universal properties of four-component fermions at large
  scattering length, we then deduce the binding energies of the three-nucleon and four-nucleon systems.
  Based on these results, we discuss the implications for primordial abundances of light nuclei, the production
  of nuclei  in stellar environments, and implications for an anthropic view of the universe.
\end{abstract}

{\flushleft\hfill{\footnotesize FTPI-Minn-20/21, UMN-TH-3922/20}}

\maketitle

\vspace{-0.2cm}

\section{Introduction}
\def\theequation{\arabic{section}.\arabic{equation}}
\label{sec:intro}

One of the most outstanding questions in physics pertains to the values of the fundamental parameters in the Standard Model. These include the gauge and Yukawa couplings, the latter being responsible for fermion masses and mixings. In the case
of the gauge couplings, some hint is available from grand unified theories where a single unified coupling is run down from a very high energy scale to the weak scale leading to predictions for the weak scale gauge couplings in reasonable agreement with experiment. The Yukawa coupling matrices are, however, a bigger mystery which includes the generation structure of fermion masses. The answer may lie in an as yet undefined future theory (e.g., a complete string theory) in which case there is hope of a 
deeper understanding. It is also possible that our Universe with its observed fundamental parameters is part of a larger structure or a {\em Multiverse}, but we have no means to know. In this case, the observed values,
may be somewhat random with no deep explanation.
However, even in that case, our specific measurements of these
parameters can not be completely random, as not all 
values will permit a Universe which supports our form of life,
which can carry out such measurements. This is often referred to as the anthropic principle.  The anthropic principle absolves
us, the  Earth dwellers, from the duty of explaining the values of the
governing constants, at least for the time  being, until data at higher
scales become  available.

The term anthropic principle was coined in  1974 by Brandon Carter \cite{carter}.
In the 1980s a few influential ``anthropic papers'' were published by Steven
Weinberg, see e.g.~\cite{Weinberg:1987dv}  (see also Refs.~\cite{Susskind:2003kw,book}).
The anthropic principle is not a predictive theory, rather it is a philosophical idea that the governing
parameters in our world should fit the intervals compatible with the existence of conscious life.
The recent LHC data show no signs to support an opposite philosophical
principle -- that of {\em naturalness}.

The most remarkable and still incomprehensible example
of anti-naturalness is the cosmological constant (for a different view,
see e.g.~\cite{Gegelia:2019fjx}). Its observed value is
suppressed by 124 orders of magnitude compared to the Planck scale $M_P^4$
(believed to be the only fundamental scale). The suppression of the
electroweak scale compared to $M_P$ is 17 orders of magnitude. The vacuum
angle $\theta$, whose natural order of magnitude $\sim 1$ is less than
$10^{-10}$ in experiment~\cite{Abel:2020gbr}. 

It is obvious that the suppression of the
cosmological constant is vital for the existence of our world.
Even if it were a few orders of magnitude larger, the Universe would have entered an inflationary stage before the onset of galaxy formation. The smallness of the $u,d$
quark masses compared to $\Lambda_{\rm QCD}$ and the fact that $m_u < m_d$
are crucial  for the genesis of heavier elements in stars. However, it is widely believed that
there are no anthropic limitations on $\theta$ and its suppression 
{\em must} be solved through a natural mechanism such as a symmetry including axions \cite{Banks:2003es,Donoghue:2003vs}.
A dedicated study of this issue \cite{Ubaldi:2008nf} revealed some $\theta$-dependence on
nuclear physics but the author concludes with the statement that ``these effects are not too
dramatic''. The authors of \cite{Banks:2003es} note with regards to the vacuum
angle $\theta$ that it ``is hard to see an anthropic argument that
$\theta$ [...] is bounded by $10^{-10}$. Moreover, in the flux vacua,
there is typically no light axion.''  For further discussions on this issue, see~\cite{Kaloper:2017fsa,Dine:2018glh}.
In the present paper we revisit this issue.

While it is certainly true (and will be made clear below) that
$\theta \sim 10^{-9}$ or
even $\theta \sim 10^{-5}$ will not change life in our world, 
it seems reasonable  to reconsider constraints imposed on 
$\theta$ from observations other than the neutron electric dipole moment (nEDM) as well as the anthropic perspective. We will see that the impact
of $\theta$ on delicate aspects of nuclear physics is similar to that of
the parameters  $|m_u|$ or $|m_d|$. Quark mass variation of nuclear
properties and reactions are considered e.g. in
Refs.~\cite{Flambaum:2002de,Beane:2002xf,Epelbaum:2002gb,Dmitriev:2002kv,Flambaum:2002wq,Dmitriev:2003qq,Jaffe:2008gd,Berengut:2009js,Bedaque:2010hr,Cheoun:2011yn,Epelbaum:2012iu,Epelbaum:2013wla,Berengut:2013nh,Lahde:2019yvr}.
Furthermore, if the variation of quark masses is due to 
an overall variation in the Yukawa couplings, it will feed into 
variations of a host of fundamental observables including the gauge couplings, and affect Big Bang Nucleosynthesis (BBN) \cite{Campbell:1994bf,Kneller:2003xf,Coc:2006sx,Ekstrom:2009ef,Coc:2012xk}, the lifetime of long-lived nuclei \cite{Olive:2002tz}, and atomic clocks \cite{Luo:2011cf}.
Strictly speaking, it would be more
appropriate to combine the absolute values of the quark masses with their
phases and analyze the limitations in the complex plane. Here, we will fix $|m_u|$ and  $|m_d|$ and let  $\theta$ vary.
Unlike Ubaldi \cite{Ubaldi:2008nf} who focused on CP-odd vertices  
and arrived at rather weak constrains, we will consider the $\theta$-dependence
due to CP-even vertices. For reviews on this and related issues,
see e.g. Refs.~\cite{Hogan:1999wh,Barnes:2011zh,Meissner:2014pma,Schellekens:2013bpa,Adams:2019kby}.

Our approach is limited in the sense that we do not vary all governing parameters simultaneously in a concerted way.
We do not explore how variations of some of them could be masked by variation of others, for instance whether the 
change of $\theta$ could be compensated  by that of $|m_{u,d}|$ or the impact of $\theta $ on, say, the vacuum energy density.
Such  a global task is a problem for the future. We will only vary $\theta$ fixing all other parameters
to their observed values.

At this point, it is worth noting that the most often discussed physical effect of $\theta$ on an observable, the nEDM arising from strong CP-violation, does not impose strong anthropic constraints on $\theta$. The nEDM stemming from the QCD $\theta$-term is \cite{Baluni:1978rf,Guo:2015tla}
\be
d_n (\bar{\theta}) = \order{10^{-16}\,\bar{\theta}\,e\, \mathrm{cm}} ,
\ee
where $\bar{\theta}=\theta + \operatorname{Arg}\det\mathcal{M}$ and $\mathcal{M}$ the quark mass matrix. Even if $\theta=\order{1}$, this is still a very small number and the physical effects of an nEDM of $\order{10^{-16}\,e\, \mathrm{cm}}$ on the evolution of the universe would still be negligible.

Note also that $\theta = \pi$ is a special point in which QCD has two
degenerate vacua, and physics changes drastically, see e.g. the lucid discussion
in Ref.~\cite{Smilga:1998dh} (and references therein). However, here we are not interested in this special point but rather in a generic situation with $0<\theta<\pi$.

As we discuss below, the value of $\theta$ does affect a host of 
hadronic properties which trigger changes in nuclear properties
such as the binding energies of nuclei. 
Changes in $\theta$ affect the pion mass which in turn alters the
neutron-proton mass difference, $\Delta m_N$ which further affects the neutron decay width. We also consider the effect of $\theta$ on multi-nucleon systems and compute changes to nuclear binding energies. 

The neutron-proton mass difference and the binding energy of deuterium, $B_d$, play a sensitive role in BBN
(see \cite{Fields:2019pfx} for the current status). As a result, changes in $\theta$ 
can substantially alter the abundances of the light elements produced in BBN. Thus we can set limits on $\theta$ (though they are weak)
entirely independent of the nEDM. However, even with large changes in
$\theta$ and large changes in the light element abundances, it is not
clear that this would cause an impediment on the formation of life in the Universe. Indeed, in a related study, 
Steigman and Scherrer~\cite{Steigman:2018wqf} addressed the question of fine-tuning in the matter-antimatter
asymmetry, as measured in terms of the baryon-to-photon asymmetry $\eta_B$. While the baryon asymmetry is reliant on the existence of CP-violation~\cite{Sakharov:1967dj}, there is no reason to suspect that the baryon asymmetry is itself related to $\theta$. 
The authors of Ref.~\cite{Kuzmin:1992up} found that even for $\theta\sim 1$ the observed  baryon asymmetry of the universe would not be altered. 
Nevertheless, changes in $\eta_B$ strongly affect the light element abundances, though it
was concluded by Steigman and Scherrer that these could not be excluded by anthropic arguments. A similar conclusion was reached in \cite{Hall:2014dfa} considering the effects of altered weak interactions on BBN.  Here, we fix $\eta_B$ and consider
the changes in abundances due changes in $\Delta m_N$ and $B_d$.

The $\theta$ induced changes
will also affect stellar evolution and can lead to very different
patterns of chemical evolution. In particular the changes in the 
nucleon-nucleon interaction, can lead to stars which yield
little or no Carbon or Oxygen, thus potentially greatly affecting the existence of life in the Universe. 

The manuscript is organized as follows: In Sect.~\ref{sec:oneN} we discuss the properties
of various mesons and the nucleons at nonzero $\theta$. First, we collect the knowledge about the
$\theta$-dependence of the corresponding hadron masses and coupling constants. Next, we focus
on the modification of the neutron-proton mass difference and the neutron decay width.
Then, we turn to the two-nucleon system in Sect.~\ref{sec:twoN}. We first construct a simple one-boson-exchange (OBE) model to describe the two-nucleon system and then display results for the deuteron, the dineutron and the
diproton with varying $\theta$. In  Sect.~\ref{sec:3and4}, we combine Wigner's SU(4) symmetry with results
from the literature to get a handle on the  $\theta$-dependence of the three- and four-nucleon systems.
Larger nuclei are briefly discussed in Sect.~\ref{sec:moreN}. Implications of these results on the nucleosynthesis
in the Big Bang and in stars are discussed in Sect.~\ref{sec:bbn} and Sect.~\ref{sec:stars}, respectively.
We end with a summary and a discussion of our anthropic view of the universe in Sect.~\ref{sec:summary}.
The appendix contains a derivation of the neutron-proton mass difference with varying $\theta$.

\section{One nucleon}
\label{sec:oneN}

In this section, we first collect the $\theta$-dependence of the various hadrons entering our study,
i.e. of the pion, the $\sigma$, $\rho$ and $\omega$ mesons as well as the nucleon mass. Our framework
is chiral perturbation theory, in which the $\theta$-dependence of the nucleon (and also of the
light nuclei) is driven by the $\theta$-dependence of the pion properties as well as the heavier
mesons, which model the intermediate and short-range part of the nucleon-nucleon interaction.
Of particular interest are the neutron-proton mass difference and the neutron decay width, which play
an important role in BBN.

\subsection{$\theta$-dependence of hadron properties}

\begin{figure}[t!]
\centering
\includegraphics[width=0.48\textwidth]{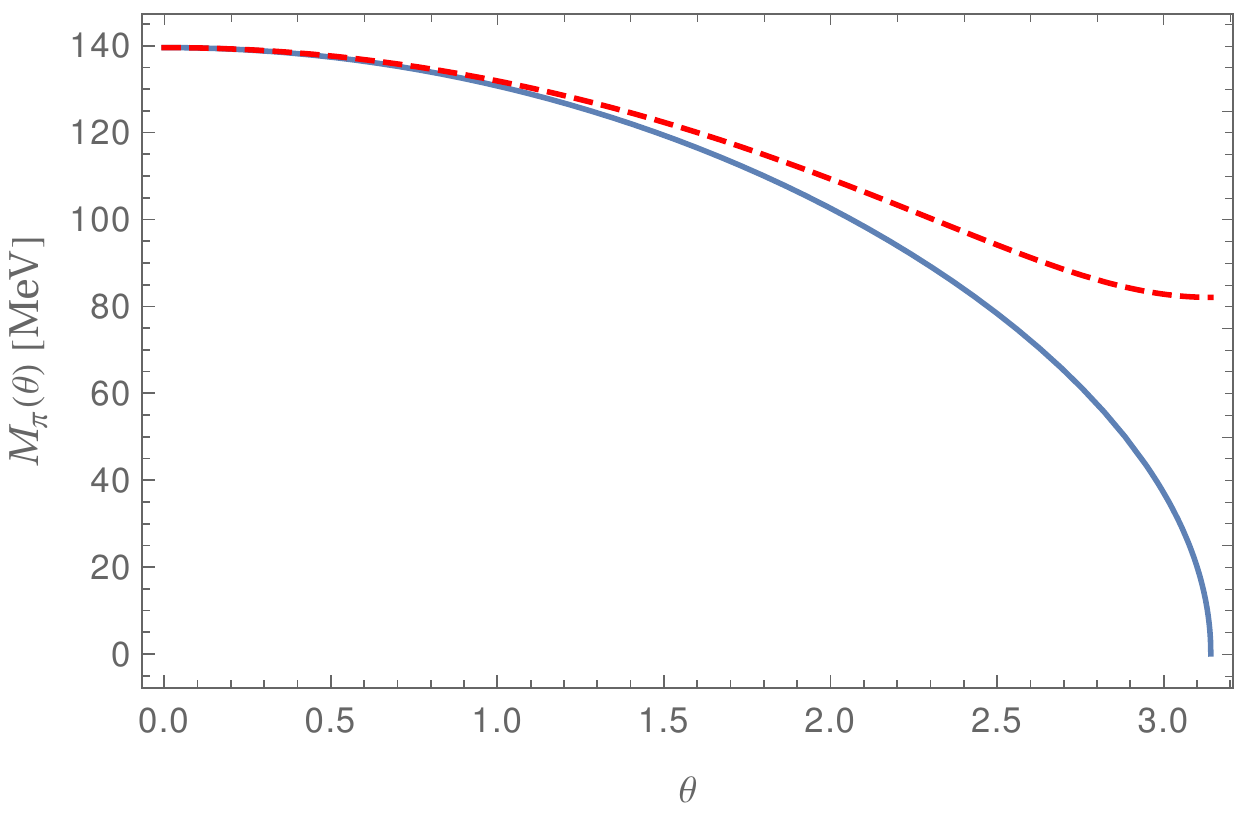}~~
\includegraphics[width=0.48\textwidth]{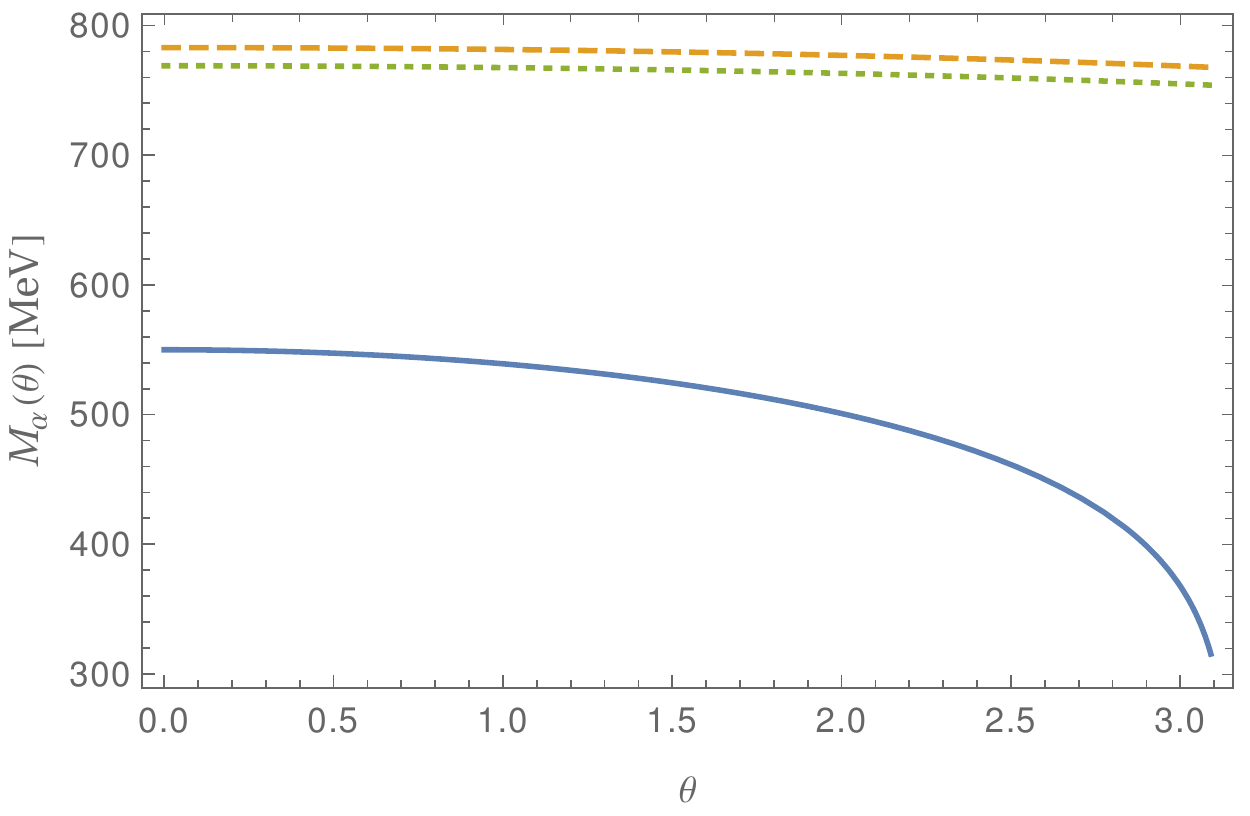}
\caption{The $\theta$-dependence of the various meson masses $M_{\alpha}$ for $\alpha=\left\{\pi,\sigma,\rho,\omega\right\}$. Left panel: The $\theta$-dependence of the pion in the case of two degenerate flavors (blue line), and in the case with $m_u\neq m_d$ (red dashed line). Right panel: The $\sigma$ meson (blue line), $\rho$ meson (green dotted line), and $\omega$ meson (orange dashed line) masses as a function of $\theta$.}
\label{fig:masses}
\end{figure}

Consider first the pion mass. We use the leading order (LO) $\theta$-dependence for two
flavors~\cite{Leutwyler:1992yt,Brower:2003yx}~\footnote{An equivalent expression for the $\theta$-dependence
of the  pion mass was also derived in a model of gluon dynamics in Ref.~\cite{Fugleberg:1998kk}.}, 
\be\label{eq:mpithetaiso}
M_\pi^2(\theta) =  {M_\pi^2}\cos\frac{\theta}{2}\sqrt{1+\varepsilon^2\tan{^2}\frac{\theta}{2}} \, ,
\ee
with $M_\pi =139.57\,$MeV, the charged pion mass, and $\varepsilon = (m_d-m_u)/(m_d+m_u)$ measures
the departure from the isospin limit. For two degenerate flavors, this reduces to
\be\label{eq:mpitheta}
M_\pi^2(\theta) =  {M_\pi^2}\cos\frac{\theta}{2} \, .
\ee
A plot of both Eq.~\eqref{eq:mpithetaiso} and \eqref{eq:mpitheta} is shown in Fig.~\ref{fig:masses} (left panel).
Since the LO contribution gives about 95\% \cite{Colangelo:2001sp}
of the pion mass at $\theta=0$, we do not need to consider higher order terms, as done e.g. in
Ref.~\cite{Acharya:2015pya}. The impact of the isospin breaking term shows up mostly as $\theta
\to \pi$. Note that while $\varepsilon\sim 1/3$, isospin symmetry is only broken by a few percent in nature
as $(m_d-m_u)/\Lambda_{\rm QCD} \ll 1$. Here, we take $m_u=2.27\,\mathrm{MeV}$ and
$m_d=4.67\,\mathrm{MeV}$ (this refers to the conventional $\overline{\rm MS}$ scheme taken
at the scale $\mu = 2\,$GeV).

The mass of the $\sigma$ as well as the masses of the $\rho$ and $\omega$ mesons when $\theta$ is varied are needed for the OBE model and are taken from Ref.~\cite{Acharya:2015pya},
assuming $M_\omega(\theta)/M_\omega(0) = M_\rho(\theta)/M_\rho(0)$ (Fig.~\ref{fig:masses}, right panel).

We consider the nucleon mass in the $\theta$ vacuum to leading one-loop order (third order in the chiral expansion),
which is given by~\cite{Brower:2003yx}~\footnote{Higher orders could be included,
  but that would go beyond the accuracy of our calculation.}
\be\label{eq:mN}
m_N (\theta) = m_0 -4c_1M_\pi^2(\theta) - \frac{3g_A^2M_\pi^3(\theta)}{32\pi F_\pi^2}~,
\ee
where $m_0 \simeq 865\,$MeV~\cite{Hoferichter:2015hva} is the nucleon mass in the chiral limit,
$g_A =1.27$ the axial-vector coupling constant, 
$F_\pi=92.2\,\mathrm{MeV}$ the pion decay constant, and $c_1 =-1.1\,$GeV$^{-1}$~\cite{Hoferichter:2015tha}
is a low-energy constant (LEC) from the second order chiral pion-nucleon Lagrangian, ${\cal L}_{\pi N}^{(2)}$,
see e.g. the review~\cite{Bernard:1995dp}. The $\theta$-dependence of the nucleon mass is
thus entirely given in terms of the pion mass, and one finds $m_N(0)=938.92\,\mathrm{MeV}$. We show the $\theta$ dependence of the nucleon mass in the left panel of Fig.~\ref{fig:nuclmasses}.

Next, we discuss the $\theta$-dependence of the coupling constants.
The $\theta$-dependence of the pion-nucleon coupling is related to
the Goldberger-Treiman discrepancy~\cite{Fettes:1998ud}
\be
g_{\pi NN}(\theta) = \frac{g_A \, m_N(\theta)}{F_\pi} \,\left(1 - \frac{2M_\pi^2(\theta)\bar{d}_{18}}{g_A} \right)~,
\ee
where $\bar{d}_{18}=-0.47\,\mathrm{GeV}^{-2}$ so that $g_{\pi NN}^2(0)/(4\pi) = 13.7$, which is in accordance with the most recent and precise value from Ref.~\cite{Baru:2011bw}.

As $g_{\rho\pi\pi}$ shows very little variation with $\theta$~\cite{Acharya:2015pya},
we can use universality relation $g_{\rho\pi\pi}= g_{\rho NN}$~\cite{Sakurai:1960ju} and keep $ g_{\rho NN}$
as well as $g_{\omega NN}$ fixed at their values at $\theta=0$ in what follows.
Matters are different for the $\sigma$. Similar to Ubaldi~\cite{Ubaldi:2008nf},
we employ the  parameterization of Refs.~\cite{Donoghue:2006rg,Damour:2007uv}.
Writing the scalar attractive piece of the nucleon-nucleon interaction as
\be
H_{\rm contact} = G_S (\bar{N}N)(\bar{N}N)~,
\ee
it is evident that
\be\label{eq:Gs}
G_S =-\frac{g_{\sigma NN}^2}{M_\sigma^2}~,
\ee
when translated to an OBE model (this corresponds to resonance saturation of the
corresponding LECs, see Ref.~\cite{Epelbaum:2001fm}). The following dependence of
$G_S(\theta)$ emerges~\cite{Ubaldi:2008nf}:
\be\label{eq:dono}
G_s(\theta)= G_S(0)\,\left(1.4 - 0.4\frac{M_\pi^2(\theta)}{M_\pi^2}\right)~,
\ee
where we have normalized again to the value at $\theta=0$.
Using Eq.~\eqref{eq:Gs} together with the known $\theta$-dependence of $M_\sigma$, we can
extract the variation of $g_{\sigma NN}$ with $\theta$. We note that the coupling $g_{\sigma\pi\pi}$
extracted from the work of Ref.~\cite{Acharya:2015pya} also decreases with $\theta$.
We now have all of the pieces of the puzzle needed to calculate the binding energies of the various
light nuclei. First, however, let us take a closer look at the neutron-proton mass
difference and the neutron decay width, which also play an important role in BBN.

\subsection{Neutron-proton mass difference}

\begin{figure}[t!]
\centering
\includegraphics[width=0.48\textwidth]{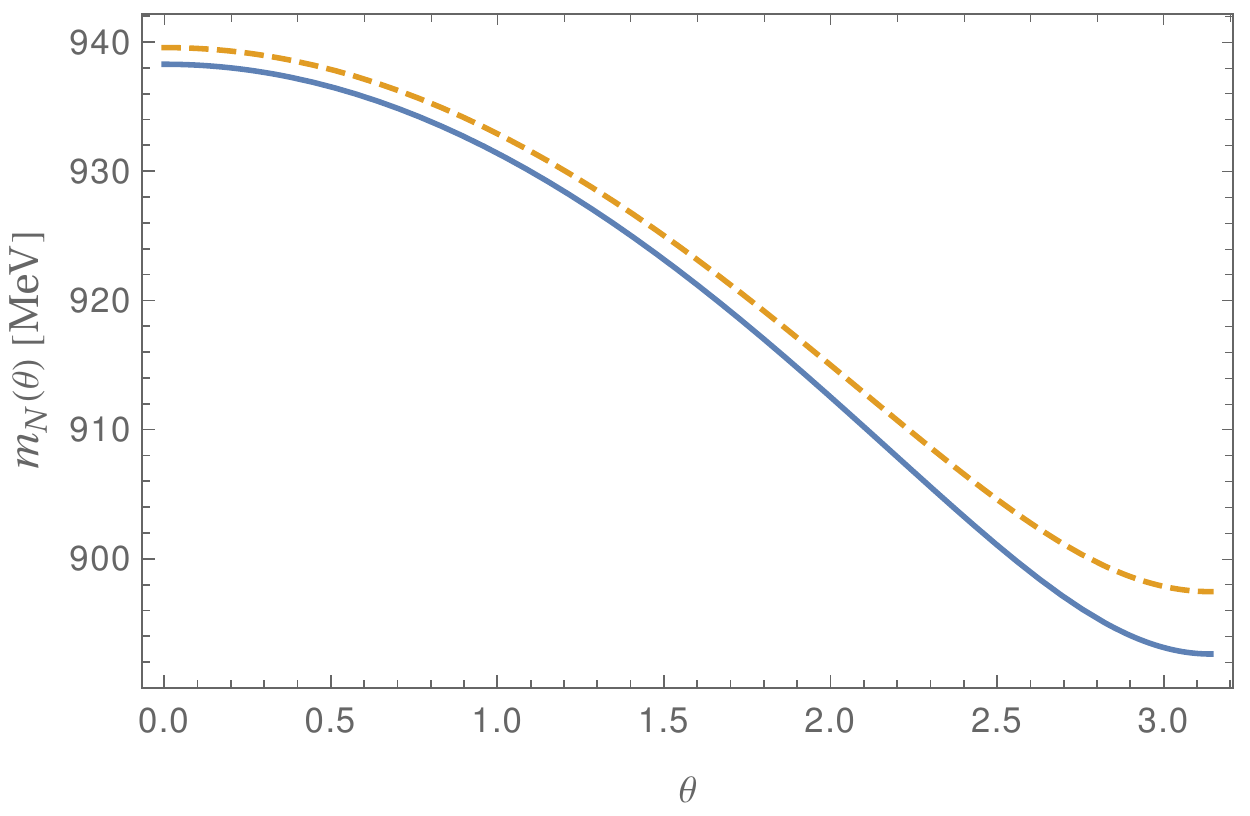}~~
\includegraphics[width=0.48\textwidth]{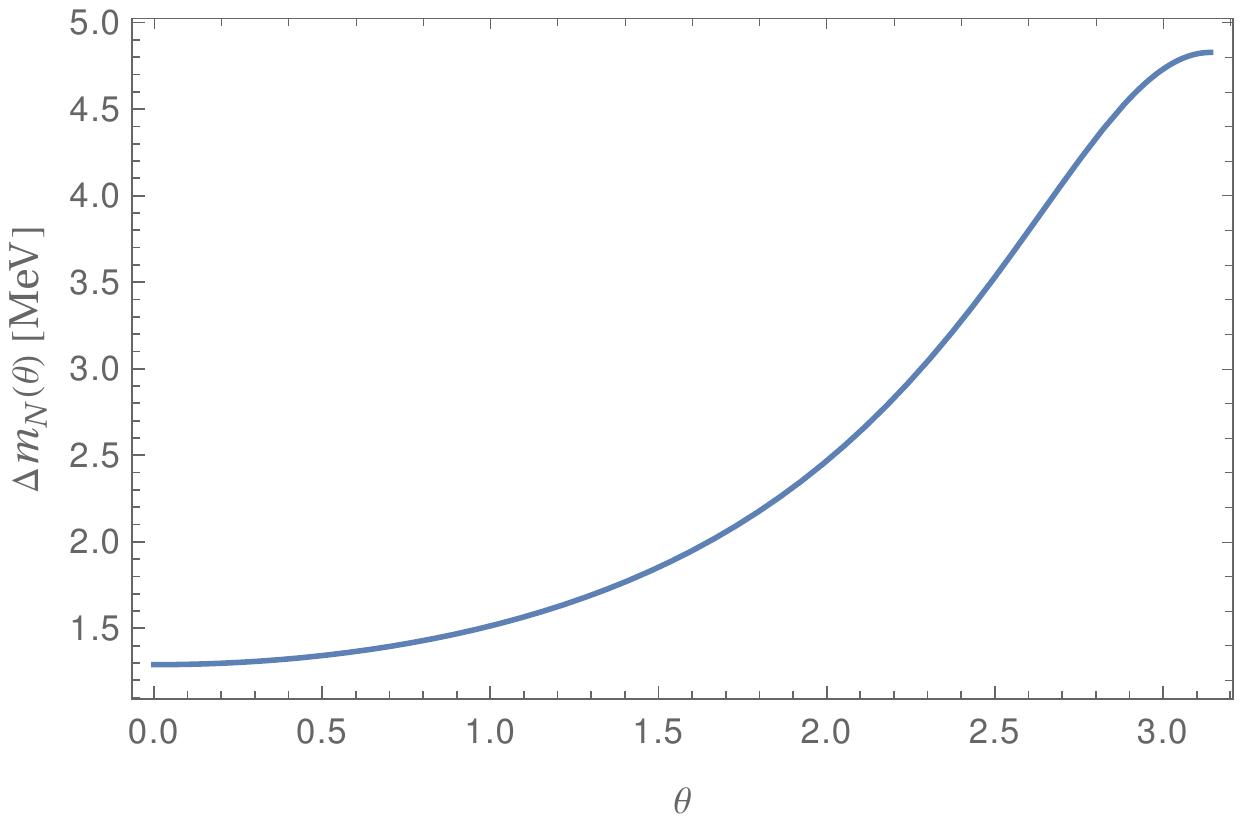}
\caption{The $\theta$-dependence of the nucleon masses $m_N$. Left panel: proton (blue line) and neutron (orange dashed line). Right panel: neutron-proton mass difference.}
\label{fig:nuclmasses}
\end{figure}

Consider the neutron-proton mass difference
\be
\Delta m_N = (m_n-m_p)^{\rm QED} + (m_n-m_p)^{\rm QCD} \simeq 1.29\,\mathrm{MeV} .
\ee
The leading contribution to the strong part to the neutron-proton mass difference arises from the second order effective pion-nucleon Lagrangian
and is given by~\cite{Bernard:1996gq}:
\be
(m_n-m_p)^{\rm QCD} = 4\,c_5\,B_0\,(m_u-m_d) +{\cal O}(M_\pi^4) = -4 \,c_5\, M_\pi^2\, \varepsilon +{\cal O}(M_\pi^4)~,
\ee
where $c_5$ is a LEC. Using the most recent determination of the electromagnetic part of this
mass difference, $(m_n-m_p)^{\rm QED} = -(0.58\pm 0.16)\,$MeV \cite{Gasser:2020mzy}, this amounts
to $(m_n-m_p)^{\rm QCD} = 1.87\mp 0.16\,$MeV and correspondingly, $c_5 = (-0.074\pm0.006)\,$GeV$^{-1}$.
In the $\theta$-vacuum, this term turns into~\cite{TV} (for a derivation, see App.~\ref{sec:appA})
\be\label{eq:deltamnp}
(m_n-m_p)^{\rm QCD} (\theta) \simeq 4 \,c_5\,  B_0\, \frac{M_\pi^2}{M_\pi^2(\theta)}\,(m_u-m_d),
\ee
i.e the strong part of the neutron-proton mass increases (in magnitude) with $\theta$, see Fig.~\ref{fig:nuclmasses} (right panel).
At $\theta \simeq 0.25$, $\Delta m_N (\theta)$ deviates already by about $1\,\%$ from its real world value, and for the range of $\theta = 1-2$, we find  $\Delta m_N (\theta) = 1.51 - 2.47\, \mathrm{MeV}$, using Eq.~\eqref{eq:mpithetaiso} for $M_\pi(\theta)$.

\subsection{Neutron decay width}

\begin{figure}[t]
\centering
\includegraphics[width=0.5\textwidth]{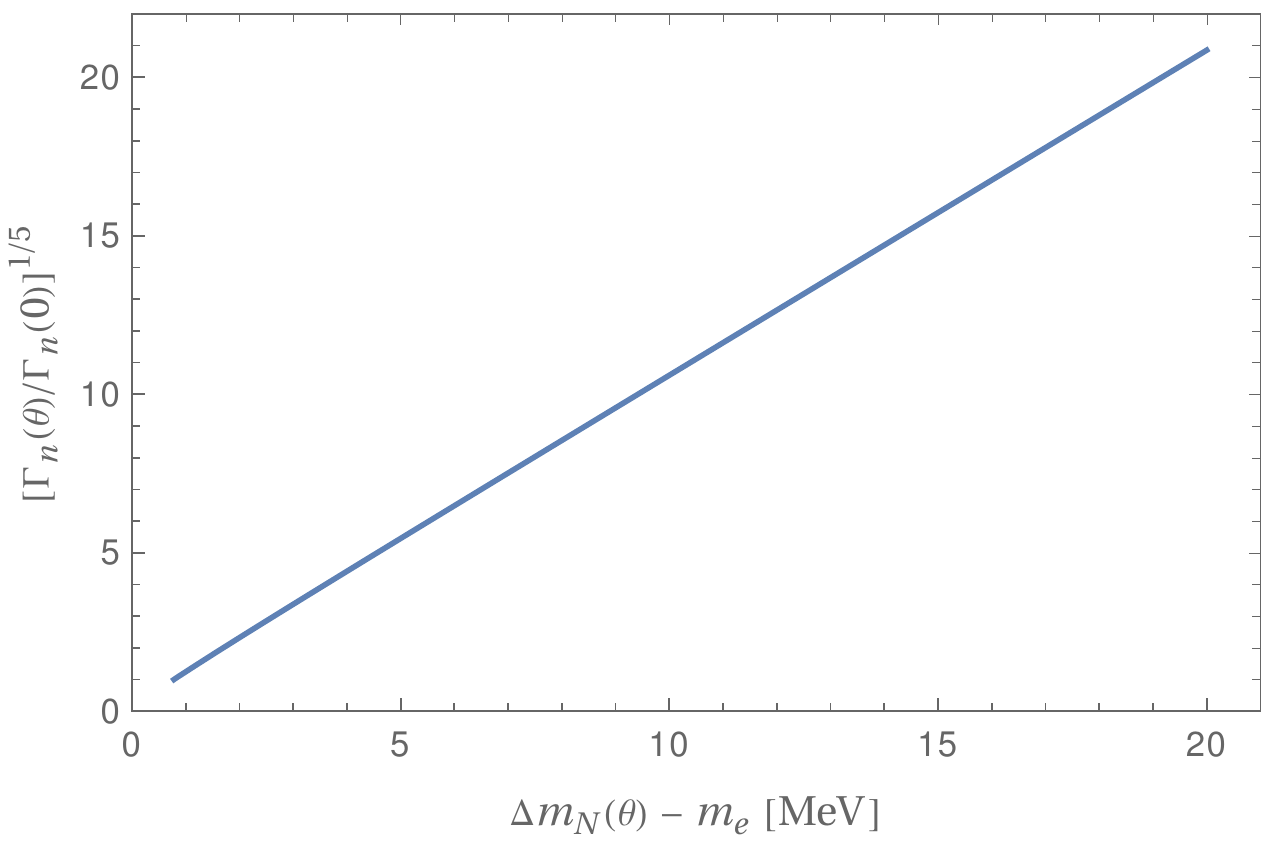}
\caption{Neutron decay width, $\Gamma_n(\theta),$ as a function of the neutron-proton mass difference.  We plot the dimensionless quantity $[\Gamma_n(\theta)/\Gamma_n(0)]^{1/5}$
  versus $\Delta m_N(\theta)-m_e$.}
\label{fig:beta_decay}
\end{figure}

\begin{figure}[t]
\centering
\includegraphics[width=0.5\textwidth]{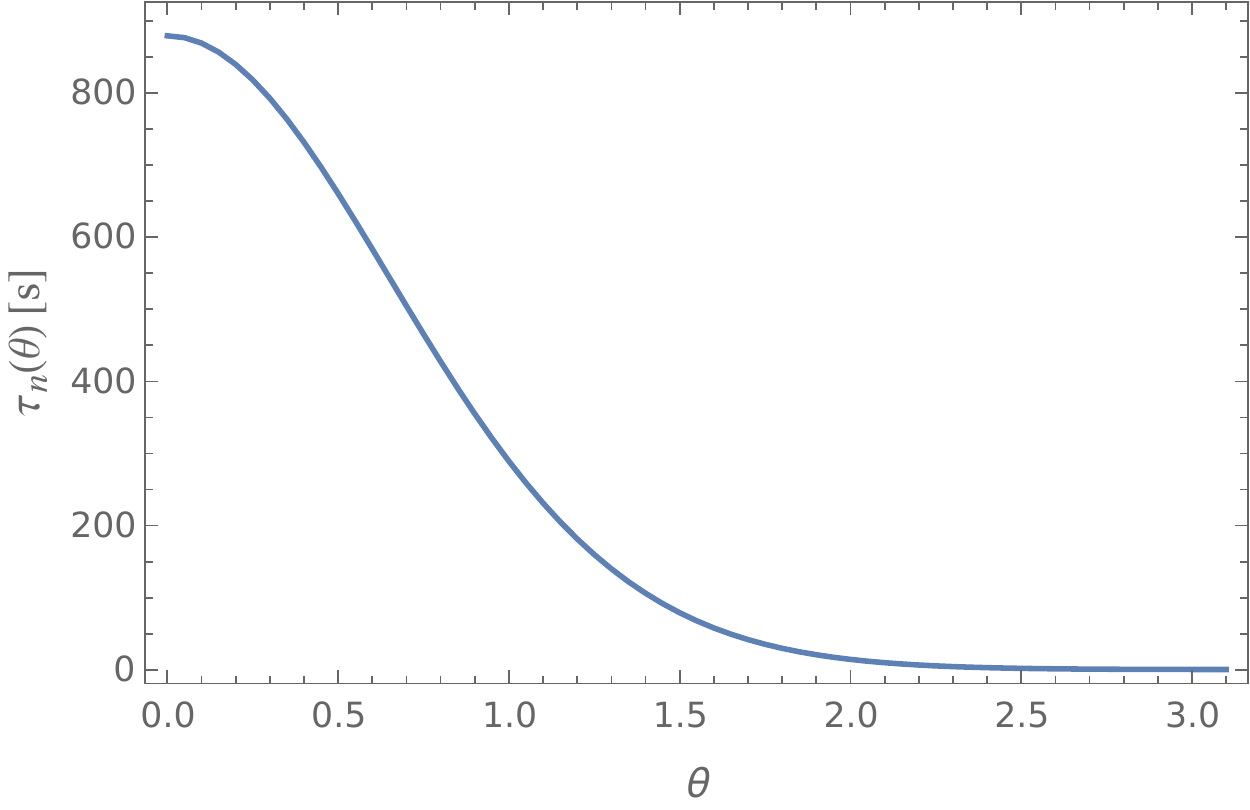}
\caption{Neutron life time, $\tau_n(\theta),$ as a function of $\theta$.  }
\label{fig:beta_decay-theta}
\end{figure}

As we increase $\theta$, the neutron-proton mass difference,  $\Delta m_N(\theta)$, becomes larger and
results in a larger three-body phase space for neutron beta decay.  This increase in the phase space integral
scales roughly as the neutron-proton mass difference to the fifth power and is dominant over any expected
$\theta$-dependence in the axial vector coupling, $g_A$.  
The neutron beta decay width can be written as (for the moment, we explicitly display factors of Planck's constant,
$\hbar$, and the speed of light, $c$, otherwise we work in natural units, $k_B =\hbar=c=1$)
\begin{equation}
\Gamma_n = \frac{m_e^5c^4}{2\pi^3\hbar^6}|{\cal M}|^2 f~,
\end{equation}
where $m_e$ is the electron mass, ${\cal M}$ is the weak matrix element and $f$ is the Fermi integral,
\begin{equation}
f = \int_0^{m_n-m_p-m_e}
F(Z,T_e)p_e T_e (m_n-m_p-m_e-T_e)^2 dT_e,
\end{equation}
where $Z=1$ is the proton charge, $T_e$ is the electron kinetic energy, $p_e$ is the electron momentum,
and $F(Z,T_e)$ is the Fermi function that takes into account Coulomb scattering \cite{Fermi:1934hr}.
In Fig.~\ref{fig:beta_decay} we plot $[\Gamma_n(\theta)/\Gamma_n(0)]^{1/5}$
versus $\Delta m_N(\theta)-m_e$ showing the linear behavior as expected. In Fig.~\ref{fig:beta_decay-theta}, the neutron mean life is shown as a function of $\theta$. We see that the lifetime drops
off very quickly when $\theta$ starts to deviate from the Standard Model value $\theta \approx 0$. As we will see this dependence plays a big role at the start of BBN. 

\section{Two nucleons}
\label{sec:twoN}

Here, we outline the formalism underlying our study of the two-nucleon system. First, we construct a simple
OBE model, that allows us to describe the binding energies of the
deuteron and the unbound dineutron and diproton  at $\theta = 0$.
Then, we discuss how these two-nucleon systems change when $\theta$ varies from $0$ to $\pi$.

\subsection{OBE model}

Consider first the case $\theta=0$. We set up an  OBE model inspired by Ref.~\cite{Ericson:1988gk}
and work with the Schr\"odinger equation, as the nucleons in the deuteron move with velocities $v \ll c$.
The corresponding OBE potential is given by
\begin{equation}
V_\mathrm{OBE} (\mathbf{q}) = \sum_{\alpha=\left\{\pi, \sigma, \omega, \rho\right\}} V_\alpha (\mathbf{q})
\end{equation}
where $\mathbf{q}$ denotes the momentum transfer. The static limit is applied, i.e.
the four-momentum transfer squared $q^2=(p^\prime-p)^2=-(\mathbf{p}^\prime-\mathbf{p})^2
=-\mathbf{q}^2$.  Setting furthermore $L=0$, i.e. focusing on the dominant S-wave and
neglecting the small D-wave contribution, the respective potentials can be reduced to
\begin{align}
V_\pi (\mathbf{q}) & = - (\boldsymbol{\tau_1} \cdot \boldsymbol{\tau_2}) ( \boldsymbol{\sigma_1}
\cdot\boldsymbol{\sigma_2}) \frac{g_{\pi NN}^2}{\mathbf{q}^2+M^2_\pi} \frac{\mathbf{q}^2}{12m_N^2}~,
\label{eq:Vp1}\\
V_\sigma (\mathbf{q},\mathbf{P}) & =  - \frac{g_{\sigma NN}^2}{\mathbf{q}^2+M^2_\sigma} \left(1
+\frac{\mathbf{q}^2}{8m_N^2}-\frac{\mathbf{P}^2}{2m_N^2}\right)~, \label{eq:Vp2}\\
V_\omega (\mathbf{q},\mathbf{P}) & = \frac{g_{\omega NN}^2}{\mathbf{q}^2+M^2_\omega}  \left(1
-\frac{\mathbf{q}^2}{2m_N^2}\left[\frac{1}{4}+\frac{1}{3}( \boldsymbol{\sigma_1}\cdot
\boldsymbol{\sigma_2})\right]+\frac{3\mathbf{P}^2}{2m_N^2}\right)~,\label{eq:Vp3}\\
V_\rho (\mathbf{q},\mathbf{P}) & =  (\boldsymbol{\tau_1} \cdot \boldsymbol{\tau_2})
\frac{g_{\rho NN}^2}{\mathbf{q}^2+M^2_\rho}\left(1-\frac{\mathbf{q}^2}{2m_N^2}\left[\frac{1}{4}
  +\frac{g_\mathrm{T}^\rho}{g_{\rho NN}}+\frac{1}{3}\left(1+\frac{g_\mathrm{T}^\rho}{g_{\rho NN}}\right)^2
  ( \boldsymbol{\sigma_1}\cdot\boldsymbol{\sigma_2})\right]+\frac{3\mathbf{P}^2}{2m_N^2}\right)  \ ,
\label{eq:Vp4}
\end{align}
where $\mathbf{P}=(\mathbf{p}^\prime+\mathbf{p})/2$. Terms $\propto (\mathbf{q}\times\mathbf{P})$,
which in coordinate space correspond to terms $\propto \mathbf{L}$, the angular momentum operator,
and terms $\propto S_{12} (\mathbf{q}) = 3 (\boldsymbol{\sigma_1}\cdot\mathbf{q})(\boldsymbol{\sigma_2}
\cdot\mathbf{q})-(\boldsymbol{\sigma_1}\cdot\boldsymbol{\sigma_2}) |\mathbf{q}|^2$, have been omitted.
The potentials depend on the total spin $S$ of the two-nucleon system through the factor $( \boldsymbol{\sigma_1}
\cdot\boldsymbol{\sigma_2})=2S(S+1)-3$ and on the total isospin $I$ through the factor
$( \boldsymbol{\tau_1}\cdot\boldsymbol{\tau_2})=2I(I+1)-3$. Note also that we omit from the start the
$\omega NN$ tensor coupling as the corresponding coupling constant $g_\omega^T$ is approximately zero, which is a good approximation, see e.g. Ref.~\cite{Ericson:1988gk,Mergell:1995bf}.

The corresponding potentials in coordinate space are of Yukawa-type and  given by
\begin{align}
V_\pi (r) & =  (\boldsymbol{\tau_1} \cdot \boldsymbol{\tau_2}) ( \boldsymbol{\sigma_1}
\cdot\boldsymbol{\sigma_2}) \frac{g_{\pi NN}^2}{4\pi} \frac{1}{12} \left(\frac{M_\pi}{m_N} \right)^2
\frac{e^{-M_\pi r}}{r}~,\label{eq:Vr1} \\
V_\sigma (r) & =  - \frac{g_{\sigma NN}^2}{4\pi} \left(1- \frac{1}{4} \left(\frac{M_\sigma}{m_N} \right)^2\right)
\frac{e^{-M_\sigma r}}{r}~, \\
V_\omega (r) & = \frac{g_{\omega NN}^2}{4\pi} \left(1 + \frac{1}{2} \left(\frac{M_\omega}{m_N} \right)^2
\left[1+\frac{1}{3}( \boldsymbol{\sigma_1}\cdot\boldsymbol{\sigma_2})\right]\right) \frac{e^{-M_\omega r}}{r}~, \\
V_\rho (r) & =   (\boldsymbol{\tau_1} \cdot \boldsymbol{\tau_2}) \frac{g_{\rho NN}^2}{4\pi}
\left(1+\frac{1}{2}\left(\frac{M_\rho}{m_N}\right)^2 \left[ 1+ \frac{g_\mathrm{T}^\rho}{g_{\rho NN}}
+\frac{1}{3}\left(1+\frac{g_\mathrm{T}^\rho}{g_{\rho NN}}\right)^2( \boldsymbol{\sigma_1}\cdot\boldsymbol{\sigma_2})
\right] \right) \frac{e^{-M_\rho r}}{r}\label{eq:Vr2} \ .
\end{align}
The OBE potential requires regularization since it is ultraviolet-divergent.
This can be most easily seen from the momentum-space representation,
Eqs.~(\ref{eq:Vp1})-(\ref{eq:Vp4}), as these potentials grow quadratically with increasing momentum transfer.
A standard regularization procedure in nuclear physics is to 
 apply either a single vertex form factor controlled by the cutoff mass $\Lambda$ for
the total  potential, or four individual form factors controlled by the cutoff masses $\Lambda_\alpha$
for each meson exchange potential. Here, we are only interested in the
binding energies of the nucleon-nucleon systems, therefore  a single form factor is sufficient.
The total OBE potential in the coordinate-space representation is then:
\begin{equation}\label{eq:VOBEreg}
V_\mathrm{OBE} (r) = \sum_{\alpha=\left\{\pi, \sigma, \omega, \rho\right\}} V_\alpha (r)
 + \frac{\Lambda}{4\pi}\frac{e^{-\Lambda r}}{r}~.
\end{equation}

At $\theta=0$, the meson masses we use are:
\be\label{eq:input}
M_\pi  = 139.57\,\mathrm{MeV}~,~~ M_\sigma  = 550\,\mathrm{MeV}~,~~ M_\omega  = 783\,\mathrm{MeV}~,~~ M_\rho  = 769\,\mathrm{MeV}~.
\ee
In order to assess the parameter dependence, we take two sets of parameters, cf. Ref.~\cite{Ericson:1988gk}:
\be\label{eq:input1}
\frac{g^2_{\sigma NN}}{4\pi}  = 14.17~,~~ \frac{g^2_{\rho NN}}{4\pi}  = 0.80~,~~
\frac{g^2_{\omega NN}}{4\pi}  = 20.0~,~~ \Lambda = 1.364~{\rm GeV}~,
\ee
which we call \emph{parameter set~I}, and 
\be\label{eq:input2}
\frac{g^2_{\sigma NN}}{4\pi}  = 8.06~,~~ \frac{g^2_{\rho NN}}{4\pi}  = 0.43~,~~
\frac{g^2_{\omega NN}}{4\pi}  = 10.6~,~~ \Lambda = 2.039~{\rm GeV}~,
\ee
which we call \emph{parameter set~II}. For both sets, we take $g_\mathrm{T}^\rho/g_{\rho NN}  = 6.1$~\cite{Ericson:1988gk,Mergell:1995bf}. After solving the radial Schr\"{o}dinger equation for the two nucleon system, one finds for both parameter sets a bound deuteron with binding energy $E_{d}=-B_{d} = -2.224\,\mathrm{MeV}$, and an unbound dineutron with $E_{nn}=-B_{nn} = 0.072\,\mathrm{MeV}$.

We now have all of the parts needed to investigate the $\theta$-dependence
of the binding energies of the various two-nucleon systems.

\subsection{Spin-triplet channel}

The bound state in the spin-triplet channel is the deuteron. Here, we work out the $\theta$-dependence
of its binding energy.

\begin{figure}[t]
\centering
\includegraphics[width=0.5\textwidth]{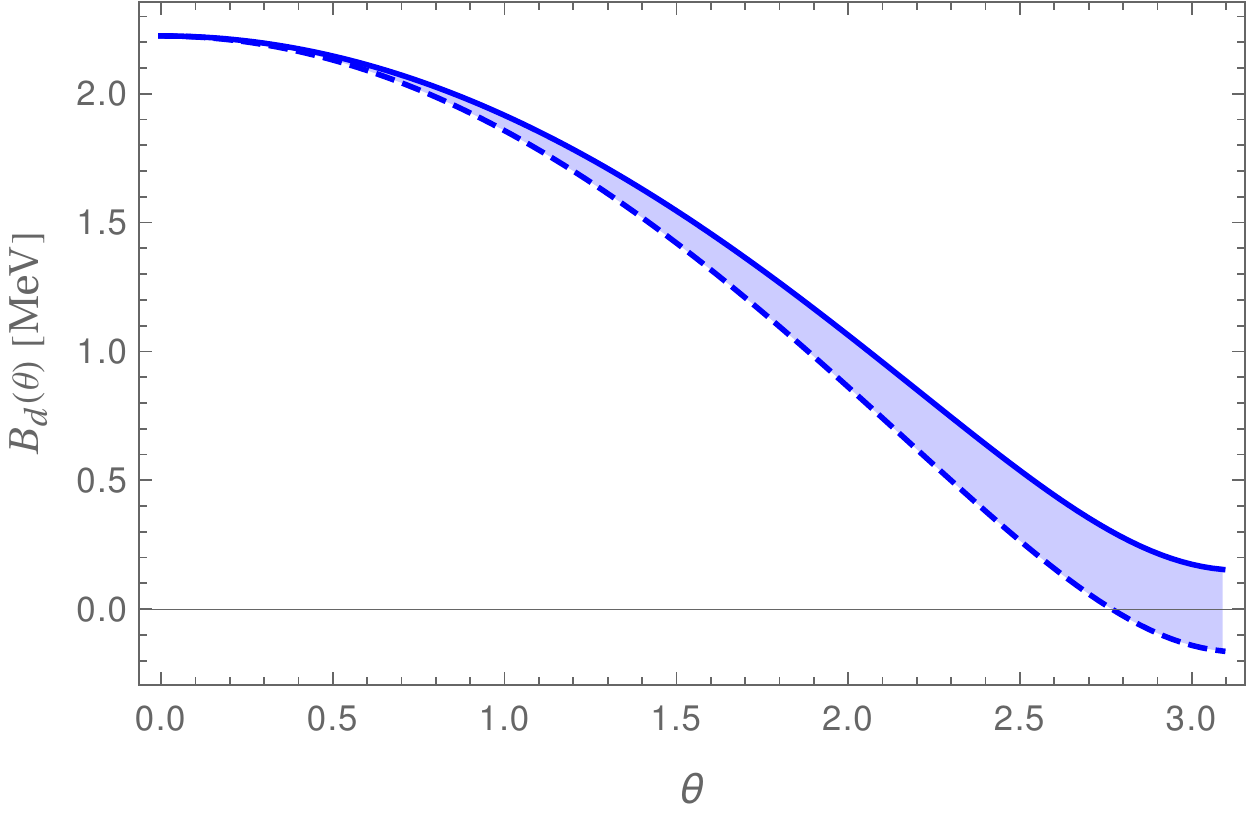}
\caption{The binding energy of the deuteron for a $\theta$-dependent OPE with all other meson couplings and
masses are kept fixed, for parameter set I, Eq.~\eqref{eq:input1} (solid line) and parameter set II, Eq.~\eqref{eq:input2} (dashed line), respectively.}
\label{fig:pionly}
\end{figure}

Consider first the  case of a $\theta$-dependent one-pion-exchange (OPE) potential, whereas all
other potentials remain constant. The resulting $\theta$-dependent
deuteron binding energy is shown in Fig.~\ref{fig:pionly}. If all OBE exchange potentials were
independent of $\theta$ except for the OPE potential, the deuteron's binding energy would
slowly decrease until the deuteron would no longer be bound for $\theta\gtrsim 2.8$ for parameter set II, Eq.~\eqref{eq:input2}.
This is the expected behavior of the
OPE potential that led to the idea that the deuteron for $\theta\neq 0$ might not be bound anymore.
This brief estimate demonstrates that the next-to-leading order contributions
calculated by Ubaldi~\cite{Ubaldi:2008nf}, which were reevaluated in~\cite{TV}, are (a) negligible (because
they are CP-odd and only account for a shift of a few percent), but also that (b) the approach of applying
first order perturbation theory is invalid, because the effects of $\theta$ on the leading order OPE potential
are not small.

However, the actual contribution of the OPE potential to the total OBE potential is very small,
which can be seen, e.g., by considering the individual potentials $V(r)$ of Eqs.~\eqref{eq:Vr1}--\eqref{eq:Vr2}.
Clearly, the smallness of the OPE contribution compared to the strong repulsion of the $\omega$ exchange
potential and the large attraction of the $\rho$ and $\sigma$ exchange suggests that, even if
the effects of $\theta$ on the scalar and vector meson masses are not as pronounced as that for the pion,
these contributions finally determine the actual $\theta$-dependence of $B_d$.

\begin{figure}[t]
\centering
\includegraphics[width=0.5\textwidth]{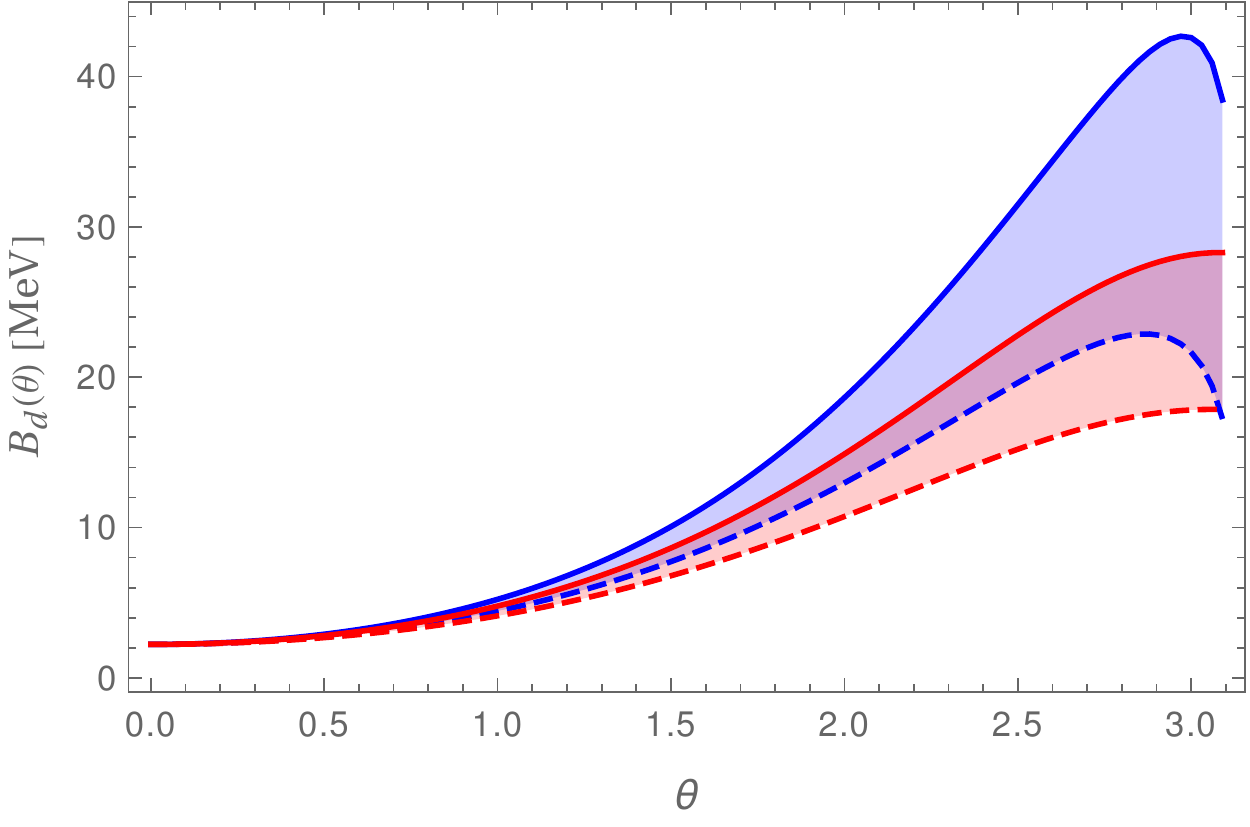}
\caption{The binding energy of the deuteron for the full $\theta$-dependent OBE model in the isospin symmetric case (blue band) and in the case of broken isospin symmetry (red band) for parameter set I, Eq.~\eqref{eq:input1} (solid lines) and parameter set II, Eq.~\eqref{eq:input2} (dashed lines), respectively.}
\label{fig:BEcouplvaried}
\end{figure}

Consider now the case of a full $\theta$-dependent OBE potential. We study two cases: first,
the isospin symmetric case with $m_u=m_d=(2.27+4.67)/2 = 3.47\,\mathrm{MeV}$, and second, the case 
of broken isospin symmetry with $m_u=2.27\,$MeV and $m_d=4.67\,\mathrm{MeV}$.
This gives the result shown in Fig.~\ref{fig:BEcouplvaried}. In the isospin symmetric case,
 we find that after increasing
and reaching a maximum at $\theta\simeq 3.0$ (parameter set I, corresponding to $B_d\simeq 42.5\,$MeV) and $\theta\simeq 2.9$ (parameter set II, corresponding to $B_d\simeq 22.8\,$MeV), respectively, the binding energy decreases and seems to approach to $B_d$ in the chiral limit, $B_d^{\rm c.l.} \simeq F_\pi^2/m \simeq 10\,$MeV~\cite{Epelbaum:2002gb}, at least in the case of parameter set II. This behavior is expected: As we have set $m_u=m_d$, $\theta\to\pi$ effectively
corresponds to $m_u=m_d\to 0$, since the charged and the neutral pion masses vanish in both cases.
Because of that, all other phenomenological quantities such as the nucleon mass and the pion-nucleon coupling
approach their respective values in the chiral limit.

In the case of broken isospin symmetry, the curve flattens and reaches its maximum as $\theta\to\pi$, which is given by $B_d\simeq 28.3\,$MeV (parameter set I) and $B_d\simeq 17.8\,$MeV (parameter set II). A useful analytic approximation for $B_d(\theta)$ is given by
\be
B_d(\theta) = 2.22 + c_1 \left(1 - \cos \theta\right) + c_2 \left( 1-\cos \theta \right)^2 + c_3 \left( 1-\cos \theta \right)^3
\ee
with
\begin{align}
\textrm{unbroken isospin symmetry:}~ & \begin{cases}
    c_1 = 9.14 \qquad c_2 = -7.19 \qquad c_3 = 6.30 \qquad {\rm Set~I}  \\    
    c_1 = 3.25 \qquad c_2 = 2.55\phantom{-} \qquad c_3 = 0.47 \qquad {\rm Set~II}     
    \end{cases} \, , \\
    \textrm{broken isospin symmetry:}~ & \begin{cases}
    c_1 = 5.68 \qquad c_2 = -1.02 \qquad c_3 = 2.36 \qquad {\rm Set~I} \\
    c_1 = 3.77 \qquad c_2 = 0.45\phantom{-} \qquad c_3 = 0.80 \qquad {\rm Set~II}     
    \end{cases}\, .
\end{align}

\subsection{Spin-singlet channel}

\begin{figure}[t]
\centering
\includegraphics[width=0.48\textwidth]{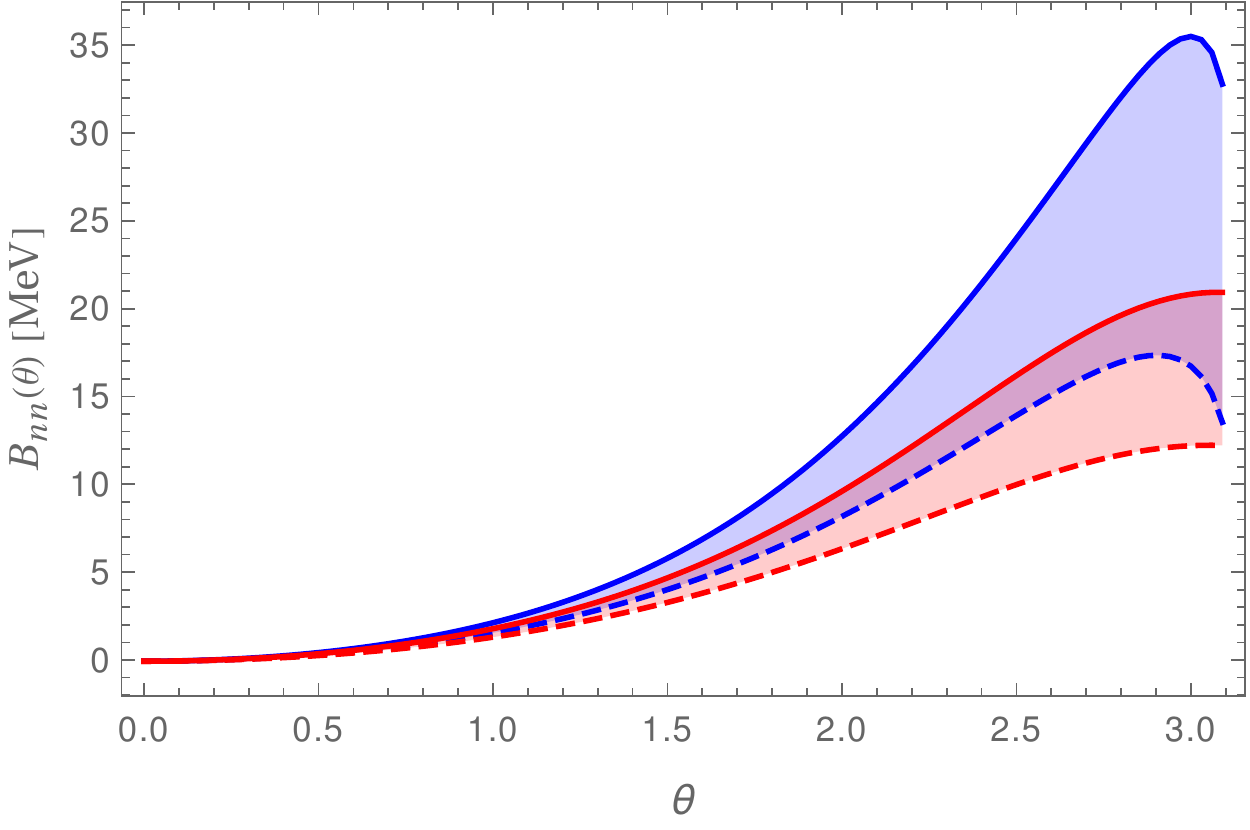}~~
\includegraphics[width=0.48\textwidth]{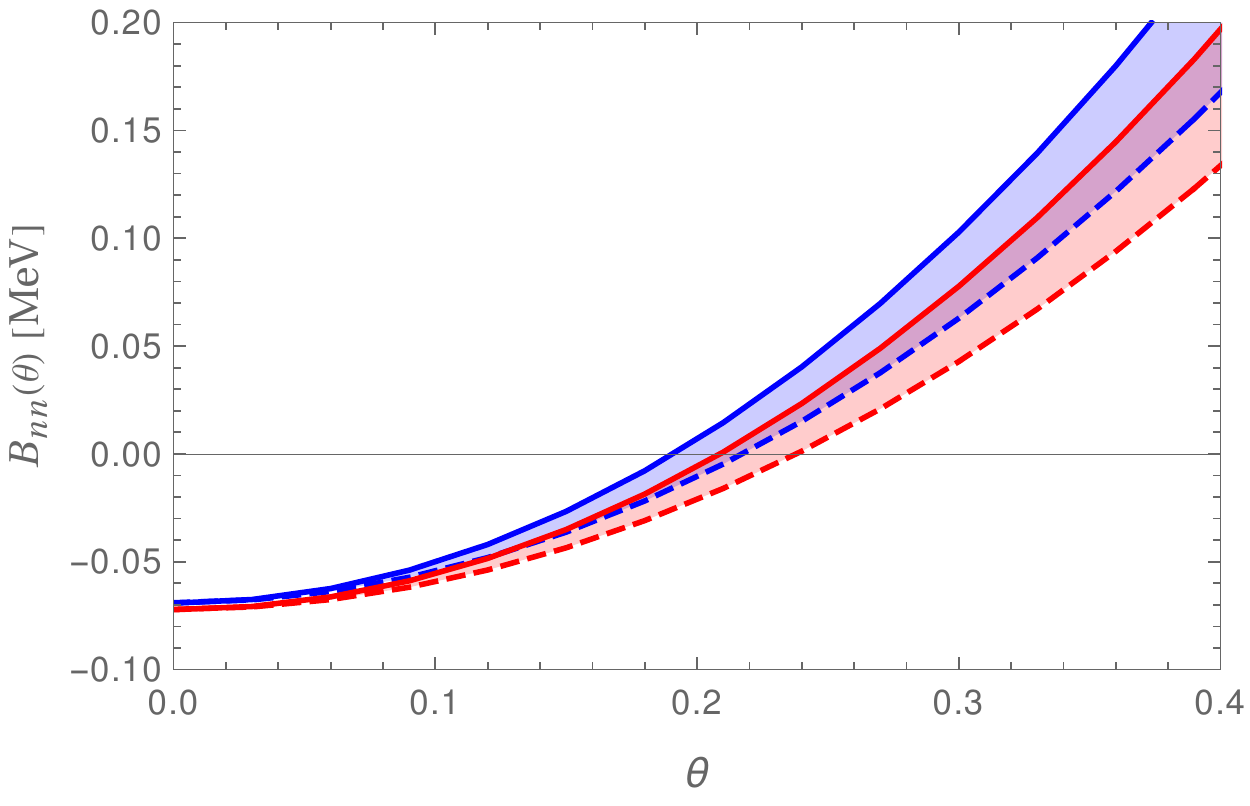}

\includegraphics[width=0.48\textwidth]{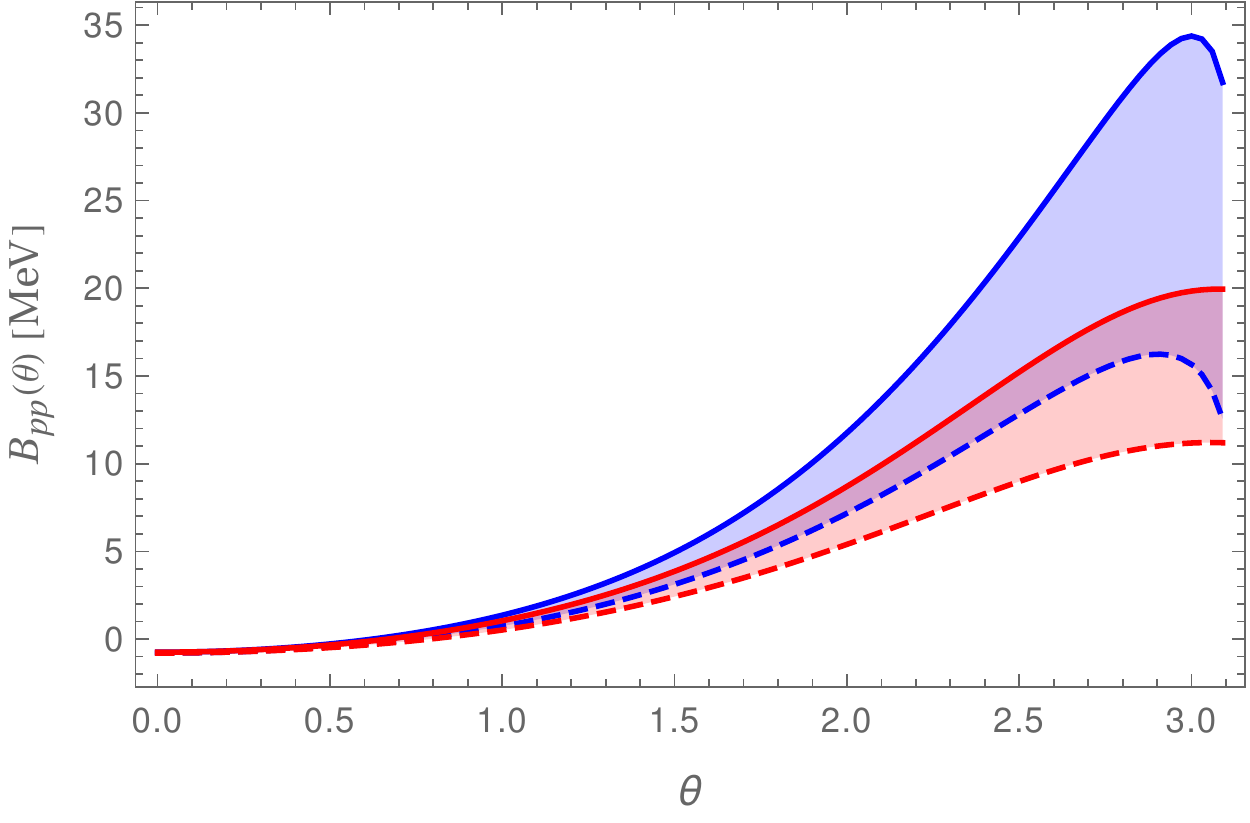}~~
\includegraphics[width=0.48\textwidth]{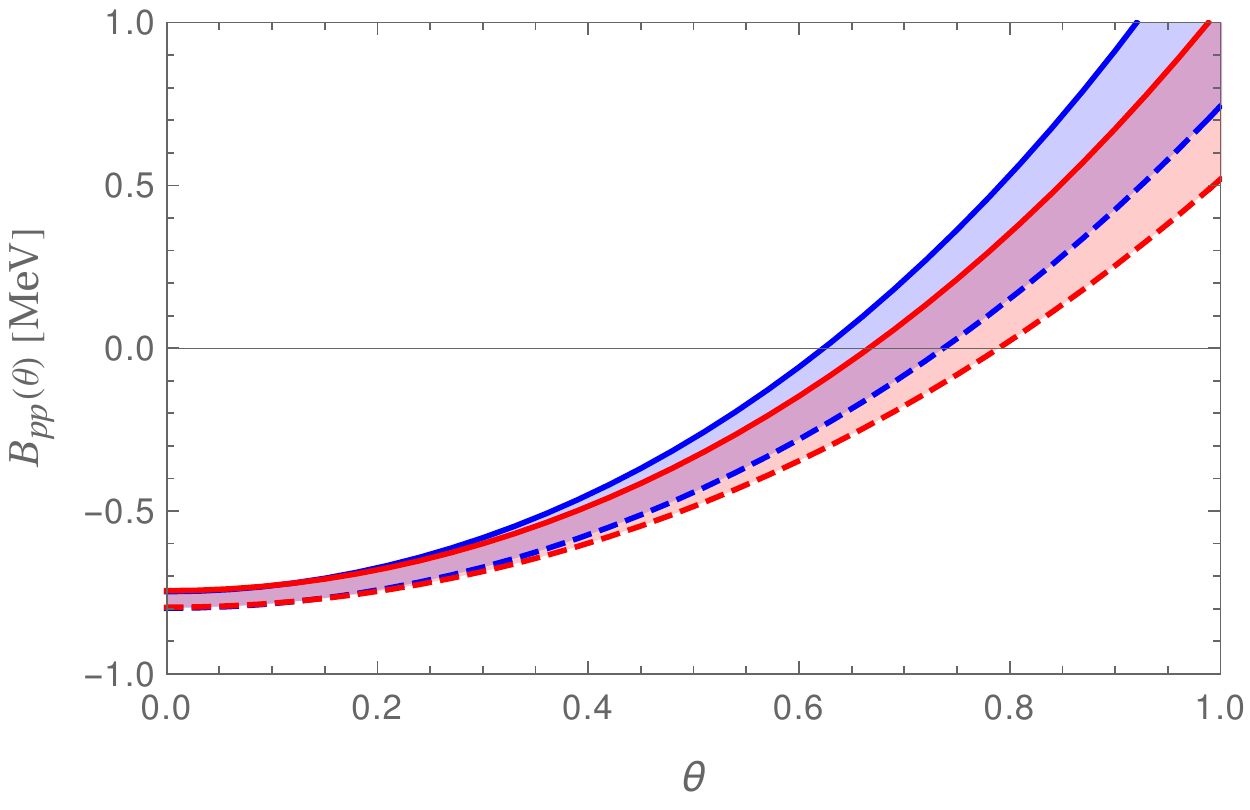}
\caption{The binding energy of the dineutron (upper panels) and of the diproton (lower panels) for the full $\theta$-dependent OBE model in the isospin symmetric case (blue band) and in the case of broken isospin symmetry (red band) for parameter set I, Eq.~\eqref{eq:input1} (solid lines) and parameter set II, Eq.~\eqref{eq:input2} (dashed lines), respectively. Left panels: Full range of $\theta$. Right panels: Zoom into the region $\theta\leq 0.4$ (dineutron), and  $\theta\leq 1$ (diproton).}
\label{fig:Bnn}
\end{figure}

The same analysis can be repeated for the dineutron with results shown in the upper panels of Fig.~\ref{fig:Bnn}. Using
Eqs.~\eqref{eq:Vr1}--\eqref{eq:Vr2}, one sees that the OPE and the $\sigma$ exchange potentials are exactly
the same for both deuteron and dineutron, i.e. with $S=1$ and $I=0$ (deuteron), and with $S=0$ and $I=1$
(dineutron). The vector exchange potentials on the other hand change in terms of the strength,
but not regarding the overall sign: the $\rho$ exchange potential is still attractive,
but weakened by about 50\,\%, whereas the $\omega$ exchange potential is still repulsive, but weakened
by about 1/3. The dineutron OBE potential is thus slightly less attractive in comparison with the
deuteron OBE potential, so the dineutron fails to be bound, as in the real world.
However, anything that happened to the deuteron OBE potential when sending $\theta\to\pi$, this also
happens to the dineutron potential, i.e. the most decisive effects come from the $\sigma$
exchange potential, which is getting stronger (while the increase of the $\rho$ exchange attraction
and the increase of the $\omega$ exchange repulsion roughly neutralize), so the dineutron becomes bound.
From the upper right panel of Fig.~\ref{fig:Bnn} one sees that this happens already for $\theta\simeq 0.18 - 0.24$.

The overall $\theta$-dependence of the dineutron's binding energy is the same as for the deuteron. Note that while the binding energy of the dineutron steadily increases, it remains smaller than the binding energy of the deuteron.

We note that a bound dineutron is also found in lattice QCD calculations with pion masses larger
than the physical one, see Refs.~\cite{Beane:2011iw,Yamazaki:2012hi,Yamazaki:2015asa,Orginos:2015aya},
which span pion masses from 300 to 510~MeV. The central binding energies in these works span the range
from 7 to 13~MeV, similar to what we find at $\theta = 1-2$.

We end with a short discussion of the diproton with $S=0$ and $I=1$. 
Referring to isospin symmetry, the only difference between the
$nn$ and the $pp$ systems is the repulsive Coulomb interaction in the latter case:
\be
V_C(r) = -\frac{e^2}{r}~,
\ee
with $e$ the elementary charge. Adding this to  our OBE potential Eq.~\eqref{eq:VOBEreg},
we find a constant shift of $-0.67$~MeV and $-0.72$~MeV for set~I and set~II, respectively,
compared to the dineutron case as shown in the lower panels of Fig.~\ref{fig:Bnn}. The only visible effect of this is that the crossover point
from the unbound to the bound case now happens at $\theta\simeq 0.6 -0.8$ (Fig.~\ref{fig:Bnn}, lower right panel).

\section{More than two nucleons}
\subsection{Three and four nucleons}
\label{sec:3and4}

We have seen that the nucleon-nucleon interaction becomes more attractive as $\theta$ increases. This is
predominantly due to the decrease in the $\sigma$ meson mass.  Since the $\sigma$ meson is a scalar particle with
zero isospin, the increased attraction is approximately the same in the spin-singlet and spin-triplet channels.
This is a realization of Wigner's SU(4) symmetry \cite{Wigner:1936dx}.  Wigner's SU(4) symmetry is an
approximate symmetry of low-energy nuclear physics where the four spin and isospin degrees of freedom are
four components of an SU(4) multiplet.  

In the SU(4) limit where the spin-singlet and spin-triplet scattering lengths are large and equal, the
properties of light nuclei with up to four nucleons follow the same universal behavior that describes
attractive bosons at large scattering length 
\cite{Bedaque:1998kg,Bedaque:1998km,Bedaque:1999ve,Platter:2004he,Platter:2004zs}.  We can use this
information to determine the $\theta$-dependent binding energies of $^3$H, $^3$He, and $^4$He.
In order to perform this analysis, we first average over nuclear states which become degenerate in the SU(4)
limit.  For the $A=2$ system, we average over the physical deuteron and spin-singlet channel to arrive at an
average binding energy of $\bar{B}_2 \simeq 1\,$MeV. For the $A=3$ system, we average over the physical $^3$H
and $^3$He systems for an average binding energy of $\bar{B}_3 = 8.1\,$MeV.  For the $A=4$ system, we take the
physical $^4$He binding energy, $\bar{B}_4 =28.3\,$MeV. 

In order to extend these binding energies to nonzero $\theta$, we use the numerical results from a study of
bosonic clusters at large scattering length~\cite{Gattobigio:2012tk}.  In particular, we use an
empirical observation from Fig.~7 of Ref.~\cite{Gattobigio:2012tk} that 
\begin{equation}
[\bar{B}_n/B]^{1/4}-[\bar{B}_2/B]^{1/4}
\end{equation}
remains approximate constant for positive scattering length $a>0$, where $B$ is a binding energy scale set by a
combination of the range of the interaction and particle mass.  Conveniently, the value of $B$ is approximately
equal to the value of $\bar{B}_4$ at infinite scattering length.  We use these empirical observations to
determine $\bar{B}_3(\theta)$ and $\bar{B}_4(\theta)$ in terms of $\bar{B}_2(\theta)$ using the approximate relation 
\begin{equation}
[\bar{B}_n(\theta)/\bar{B}_4(0)]^{1/4}-[\bar{B}_2(\theta)/\bar{B}_4(0)]^{1/4}=
[\bar{B}_n(0)/\bar{B}_4(0)]^{1/4}-[\bar{B}_2(0)/\bar{B}_4(0)]^{1/4}~.
\end{equation}

\begin{figure}[t]
\centering
\includegraphics[width=0.48\textwidth]{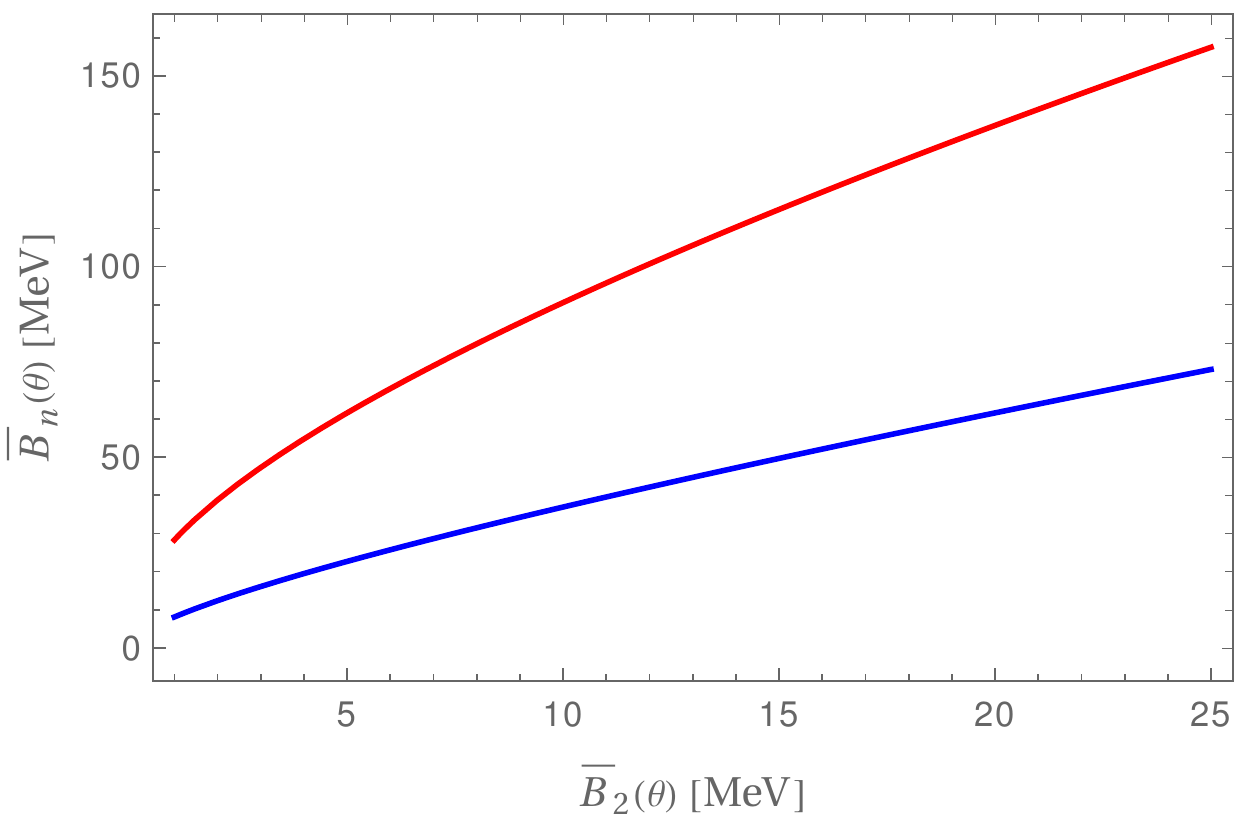}~~
\includegraphics[width=0.48\textwidth]{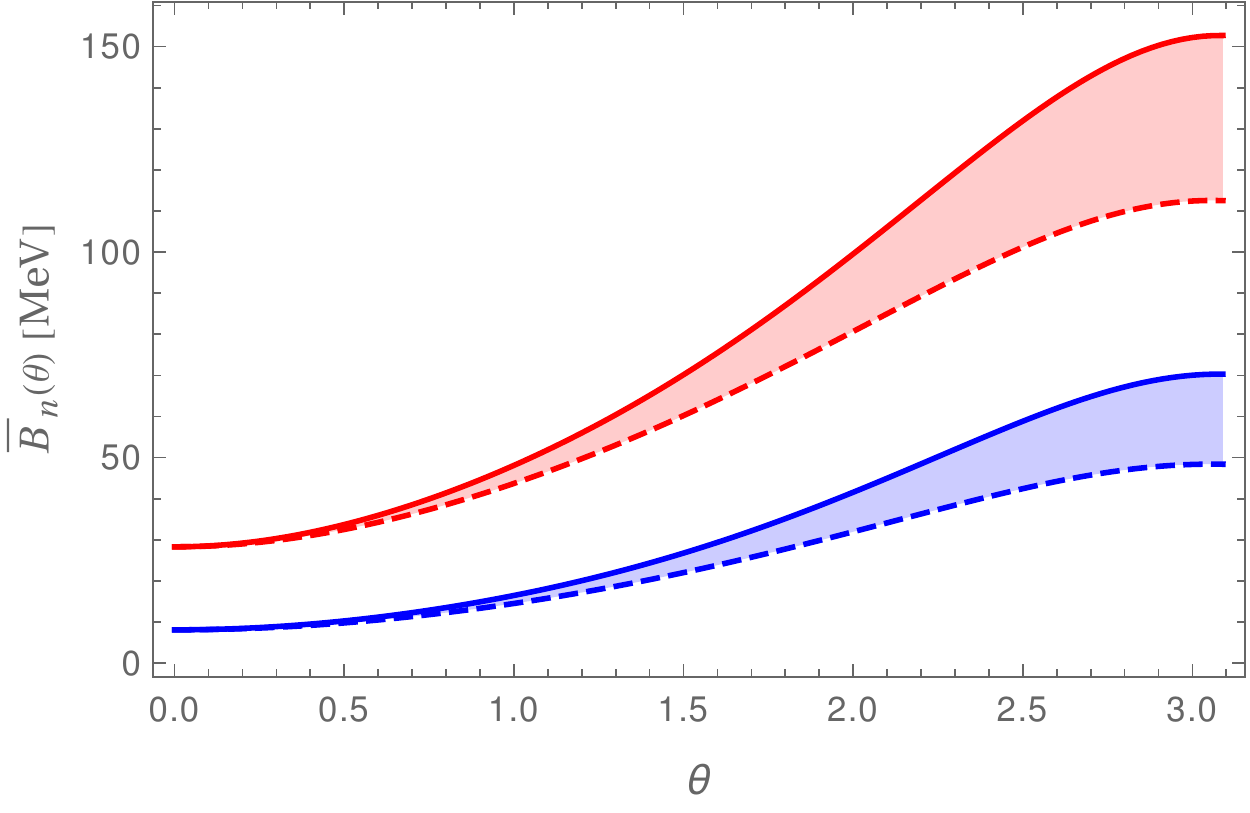}
\caption{Left panel: The SU(4)-averaged binding energy of the three- and four-nucleon systems, $\bar{B}_3(\theta)$ (blue) and
  $\bar{B}_4(\theta)$ (red), versus the SU(4)-averaged binding energy of the two-nucleon system, $\bar{B}_2(\theta)$. Right Panel: $\bar{B}_3(\theta)$ (blue band) and
  $\bar{B}_4(\theta)$ (red band) versus $\theta$, taking isospin breaking effects into account, for parameter set I (solid lines), and parameter set II (dashed lines).}
\label{fig:A34}
\end{figure}

In Fig.~\ref{fig:A34} we show the SU(4)-averaged binding energy of the three- and four-nucleon systems,
$\bar{B}_3(\theta)$ and $\bar{B}_4(\theta)$, versus the SU(4)-averaged binding energy of the two-nucleon
system, $\bar{B}_2(\theta)$, in the left panel and directly as a function of $\theta$ in the right panel.  Our results are similar to those obtained in Ref.~\cite{Barnea:2013uqa},
which were computed using hyperspherical harmonics and auxiliary-field diffusion Monte Carlo.

\subsection{More than four nucleons}
\label{sec:moreN}

In Ref.~\cite{Elhatisari:2016owd} the authors noted that the strength of the $^4$He-$^4$He interaction is
controlled by the strength and range of the SU(4)-invariant local nucleon-nucleon interaction. By local we mean
an interaction that is velocity independent. We have noted that as  $\theta$ increases, the range and strength of the SU(4)-invariant local nucleon-nucleon interaction increases due to the $\sigma$ exchange contribution. We have already observed the increase in the binding energies of the two-, three-, and four-nucleon systems.  As discussed in Ref.~\cite{Elhatisari:2016owd}, the increase in
the range of the local interaction will also cause alpha-like nuclei to become more bound.  This is discussed further in Sec.~\ref{sec:bbn} and Sec.~\ref{sec:VIB}.

Across the nuclear chart, the binding energy per nucleon will increase with $\theta$, and the relative
importance of the Coulomb interaction will decrease. As a result, the density of nucleons at nuclear
saturation will also rise. Given the increase in the neutron-proton mass difference and decreased importance of the
Coulomb interaction, the line of nuclear stability will shift towards nuclei with equal numbers of neutrons
and protons and extend to larger nuclei.

\section{Big Bang nucleosynthesis}
\label{sec:bbn}

In the early universe the temperature, $T$, is high enough to keep neutrons
and protons in thermal equilibrium through the weak interactions
\begin{align}
n + e^+ &\leftrightarrow p + \bar{\nu}_e ~,  \nonumber \\
n + \nu_e &\leftrightarrow p + e^-~,  \nonumber \\
n & \leftrightarrow p + e^- + \bar{\nu}_e~.
\label{beta}
\end{align}
The weak interaction rates scale as $T^5$ and can be compared with the expansion rate of the Universe, given by the Hubble parameter, $H \propto T^2$ in a radiation dominated Universe. 
As the temperature drops, the weak rates freeze-out, i.e. they fall out of equilibrium when they drop below the Hubble rate. In standard BBN, this occurs at a temperature, $T_f \simeq 0.84$ MeV. In equilibrium,
the ratio of the number densities of neutrons to protons
follow the Boltzmann distribution 
\begin{equation}
\frac{n}{p} \equiv \frac{n_n}{n_p}\simeq\exp\left[-\frac{{ \Delta m_N}}{{T}}\right]~.
\label{npratio}
\end{equation} 
At freeze-out, this ratio is about 1/4.7. 
The neutron-to-proton ratio is particularly important,
as it is the primary factor determining the $^4$He abundance.
The $^4$He mass fraction, $Y$, can be written as
\be
Y = 2 X_n \equiv \frac{2(n/p)}{1+(n/p)} \, ,
\ee
and its observed value is $Y = 0.2449 \pm 0.0040$ \cite{Aver:2015iza}.
Further, $X_n$ is the neutron fraction.
A change in $\theta$, will therefore invariable affect the $^4$He
abundance, primarily through the change in $\Delta m_N$. 
 While the change in $\theta$ and $\Delta m_N$ does induce a change in $T_f$, this is minor ($<$ 10\% in $T_f$)
and we neglect it here.

The helium abundance, however, is not determined by $(n/p)$ at freeze-out, but rather by the ratio at the time BBN begins. 
At the onset of BBN,  deuterons are
produced in the forward reaction  
\begin{equation}
  n + p \leftrightarrow d + \gamma\, .
\end{equation}
However, initially (even though $T < B_d$), deuterium is
photo-disintegrated by the backward reaction at temperatures
$T_d \gtrsim 0.1$ MeV. This delay, often called the deuterium bottleneck, is caused by the large excess of photons-to-baryons (or the smallness of $\eta_B$), and allows time
for some fraction of the free neutrons to decay. A rough estimate
of the temperature at which deuterium starts to form is 
\be
T_d \sim - \frac{B_d(\theta)}{\ln \eta_B}
\label{TBD}
\ee
which for $\theta=0$ yields $T_d \sim 0.1\,$MeV. A more accurate
evaluation would find $T_d \approx 0.064$ MeV. 
Below this temperature, 
the photo-disintegration processes become negligible and nucleosynthesis begins.

\begin{figure}[t]
\centering
\includegraphics[width=0.5\textwidth]{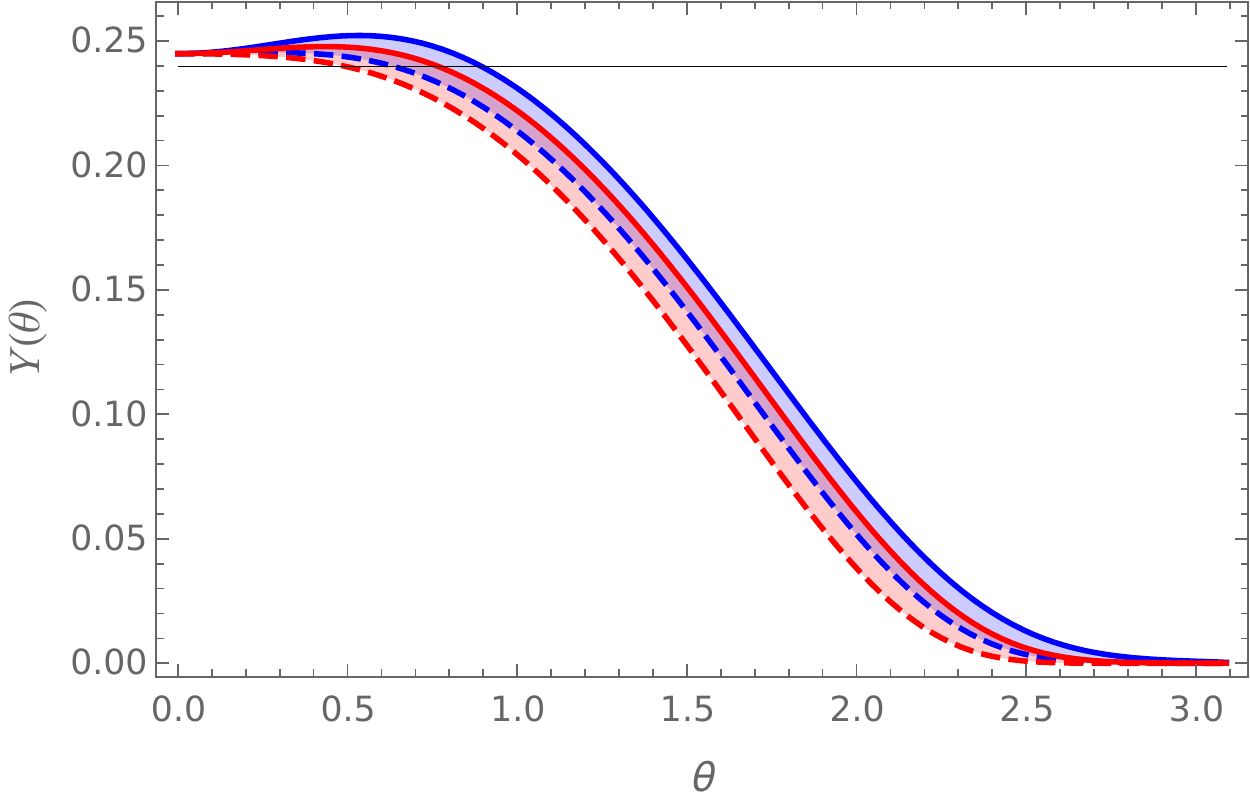}
\caption{ The Helium mass fraction, $Y$, as a function of $\theta$ in the isospin symmetric case (blue band) and in the case of broken isospin symmetry (red band) for parameter set I, Eq.~\eqref{eq:input1} (solid lines) and parameter set II, Eq.~\eqref{eq:input2} (dashed lines).}
\label{fig:helium}
\end{figure}

A change in the starting time of BBN changes the $(n/p)$ at freeze-out
or more accurately the neutron fraction, $X_n$, at freeze-out by
\be
X_n ({T_d}) = X_n (T_f) e^{-t_d/\tau_n}
\label{Gammat}
\ee
where $t_d$ is the age of the Universe corresponding to the temperature, $T_d$. As noted earlier, $\Gamma_n \propto (\Delta m_N)^5$, and in a radiation dominated Universe, $t \propto T^{-2}$, so that from (\ref{TBD}), $t_d \propto B_d^{-2}$. 
Thus using the dependencies of $\Delta m_N$, $\tau_n$, and $B_d$
on $\theta$, we can calculate $Y(\theta)$ as shown in Fig.~\ref{fig:helium}. Note that to produce Fig.~\ref{fig:helium} we have used the numerical values of
$\Gamma_n$ and $B_d$ as in Figs.~\ref{fig:beta_decay-theta} and \ref{fig:BEcouplvaried}, rather than the analytic approximations.

As one can see in the figure, the Helium mass fraction is relatively
flat for $\theta \lesssim 1$.  This is due to competing effects 
in determining $Y$. As we saw in the right panel of Fig.~\ref{fig:nuclmasses}, the neutron-proton mass difference increases with $\theta$. This strongly suppresses the neutron-to-proton ratio, as seen in Eq.~(\ref{npratio}).
Furthermore, because $\Gamma_n \propto (\Delta m_N)^5$, 
an even stronger suppression in $Y$ occurs due to the increased neutron decay rate as seen in Eq.~(\ref{Gammat}). However these decreases are largely canceled at low $\theta$ by the increase in $B_d$, which causes BBN to begin earlier, leaving less time for neutron decay. In fact, for set I parameters, at low $\theta$ this is the dominant change in $Y$ and causes an increase in the Helium abundance. The maxima occur at $\theta = 0.42 (0.54)$ and $Y = 0.248 (0.252)$ for broken (unbroken) isospin symmetry. 
Requiring $Y > 0.24$, sets upper limits on $\theta$
of roughly 0.77 (0.50) for broken isospin, and 0.89 (0.61) for unbroken 
isospin for parameter Sets I (II), respectively. For larger values of $\theta$, the Helium abundance will drop below the observationally
inferred limit,\footnote{While we have not run a nucleosynthetic chain in a numerical BBN analysis, the analytic approximation for $Y$ is quite good.
For $\theta = 0$, we have $Y = 0.2467$,
while the current result from a full BBN analysis is $Y= 0.24696$ \cite{Fields:2019pfx}.} however, as we note earlier, 
it is not clear that a Universe with primordial Helium and $Y < 0.05$
would prevent the formation of life and therefore can not be excluded anthropically. We also note that an increase in $\theta$ and an increased $B_d$ will lead to an increase in the BBN value for D/H \cite{Coc:2012xk}
which is now very tightly constrained by observation D/H $= 2.53 \pm 0.03$ \cite{CPS}.

An interesting subtlety occurs in the case of unbroken isospin symmetry for parameter Set I. As one can see in Fig.~\ref{fig:BEcouplvaried}, the deuteron binding energy increases above $\sim 30$ MeV, when $\theta \gtrsim 2.4$. In this case, there
is effectively no deuterium bottleneck, as the backward reaction in (\ref{TBD}) shuts off before weak decoupling. The Helium abundance, however, is highly suppressed due to the large value of $\Delta m_N \gtrsim 3.5$ MeV and $Y \lesssim 0.05$.

As described above, the other two potentially bound dimers, the dineutron and the diproton, become bound at $\theta\simeq 0.2$ and  $\theta\simeq 0.7$, respectively.
 Variations in the binding energy of the dineutron is expected to have little
effect on the primordial abundances provided its absolute value remains smaller than the deuteron's binding energy \cite{Kneller:2003ka,Coc:2006sx,MacDonald:2009vk}.
Considering that, in this work the variations on the binding energy of
the deuteron are only of a few percent, we do not expect any important role played by the binding energy of the dineutron in the calculations. For large $\theta$, although diprotons are bound, their binding energy remains below that of Deuterium and it was argued that diproton production freezes-out before the diproton bottleneck is broken \cite{bradford,MacDonald:2009vk}.

Before concluding this section, we consider the possible impact of changes in the binding energy of unstable nuclei. 
In \cite{Ekstrom:2009ef}, changes in the nuclear part of the nucleon-nucleon potential were parameterized as 
\be
V_N({\bf r}_{ij}) = (1 + \delta_{NN})V_N^0({\bf r}_{ij}) \, ,
\ee
where $V_N^0({\bf r}_{ij})$ is the nucleon-nucleon potential based 
on the Minnesota force adapted to low mass systems \cite{Thompson:1977zz}. The binding energy of $^8$Be, 
was found to be \cite{Ekstrom:2009ef}
\be
B_8 = (-0.09184 + 12.208 \delta_{NN})~{\rm MeV}
\label{be8bind}
\ee
indicating that $^8$Be becomes bound when $\delta_{NN} \ge 0.00752$.
\footnote{$^5$He and $^5$Li are unbound by 0.798 and 1.69 MeV, respectively, i.e. roughly an order of magnitude more than $^5$Be
requiring a very substantial change in $\delta_{NN}$ and we do not 
consider this possibility here.}
The binding energy of Deuterium is also affected by a change in the 
nucleon-nucleon potential
\be
B_d(\theta) = \left(1+5.716\,\delta_{NN} (\theta) \right) B_d(0) ,
\ee
where we have implicitly here made $\theta$ the origin of this change. From these expressions, we estimate that $^8$Be becomes
bound when $B_d(\theta) = 2.32$ MeV or when $\theta$ is 0.21 (0.23) for broken isospin, and 0.19 (0.22) for unbroken 
isospin for parameter Sets I (II), respectively.

For stable  $^8$Be, it may be possible in principle that BBN
produce elements beyond $^7$Li. As we discuss further in the next section, changes in the nuclear potential
strongly affects the triple $\alpha$ process and the production of 
Carbon and Oxygen in stars \cite{Ekstrom:2009ef}. In the context of BBN, 
stable  $^8$Be increases the importance of two reactions
$^4$He$(\alpha,\gamma){}^8$Be and $^8$Be$(\alpha,\gamma){}^{12}$C.
Nevertheless, the detailed study in \cite{Coc:2012xk}, found that while some $^8$Be is produced in BBN (with a mass fraction of $10^{-16}$ for $\delta_{NN} = 0.0116$), no enhancement of Carbon occurs as the temperature and density in the BBN environment is substantially below that in stars and the production rates are
inefficient.

\section{Stellar nucleosynthesis}
\label{sec:stars}

\subsection{Hydrogen burning}

The effects of nonzero $\theta$ will also be manifest in 
stellar nucleosynthesis.  We first consider main sequence stars undergoing hydrogen burning.  The first step of hydrogen burning is proton-proton fusion,
\begin{equation}
 p + p \rightarrow d + e^+ + \nu_e~.
\end{equation}
For $\theta \lesssim 0.5$ proton-proton fusion is not significantly altered from how it occurs in the physical Universe.  However for $\theta \gtrsim 0.7$, the diproton becomes bound and the first step in hydrogen burning
can proceed many orders of magnitude faster via radiative capture, 
\begin{equation}
 p + p \rightarrow pp + \gamma~.
\end{equation}
The diproton can then subsequently decay via the weak interactions to a deuteron,
\begin{equation}
pp \rightarrow d + e^+ + \nu_e~.
\end{equation}
We note that while the neutron-proton mass difference grows with $\theta$, the diproton still has a higher mass than the deuteron due to the larger binding energy of the deuteron.

Initially it was thought the rapid processing of protons
to diprotons would lead to stars with extremely short lifetimes,
so short so as to prevent the evolution of life on planets.
However, stellar structure compensates, and burning occurs 
at lower temperatures and densities \cite{bradford,Barnes:2015mac} and though the stars would be
different, it is not clear that there is an anthropic argument
against such stars.

\subsection{Constraints on {\boldmath $\theta$} from the anthropic principle}
\label{sec:VIB}

The anthropic principle can constrain $\theta$ if changes in $\theta$
result in a departure of normal stellar evolution so great that 
planetary life would not occur. 
Therefore, we could at minimum require that (a) enough metals (in the astronomical sense) are available, and that (b) the lifetime of stars with higher metallicity (thus allowing for rocky planets with potentially living beings) is long enough that intelligent life can evolve. Perhaps two of the most important elements for the production of life as we know it are Carbon and Oxygen.

There have been many studies relating the sensitivity of Carbon production to fundamental physics in relation to the anthropic principle \cite{livio,Fairbairn:1999js,Csoto:2000iw,Oberhummer:2000zj,Schlattl:2003dy,Tur:2008uw,Epelbaum:2012iu,Epelbaum:2013wla,Lahde:2019yvr,Huang:2018kok}. The production of ${}^{12}{\rm C}$ in stars requires a triple fine tuning: (i) the decay lifetime of ${}^8{\rm Be}$, is relatively long, and is of order $10^{-16}$~s, which is four orders of magnitude longer than the scattering time for two $\alpha$ particles, (ii) there must exist an excited state of Carbon which lies just above the energy of ${}^8{\rm Be}+\alpha$ and (iii) the energy level of ${}^{16}{\rm O}$ which sits at 7.1197~MeV must be non-resonant and below the energy of ${}^{12}{\rm C}+\alpha$, at 7.1616~MeV, so that most of the produced Carbon is not destroyed by further stellar processing. 
It is well known of course, that the existence of the excited state of $^{12}$C was predicted by Hoyle \cite{Hoyle:1954zz}. Any change in fundamental physics which affects the position of the Hoyle resonance, could severely affect the production of Carbon and Oxygen and ultimately the existence of life.

We saw that it is perhaps not possible to place anthropic bounds on $\theta$ from BBN, as it is hard to see why a universe with
a paucity of Helium would prevent star formation or stellar processing. It is however possible to set some constraints
on $\theta$ based on its effect on the triple $\alpha$ process leading to Carbon production in stars. 
In addition to the change in the $^8$Be binding energy given in
Eq.~(\ref{be8bind}), changes in $\theta$ and thus changes in the nucleon-nucleon 
potential, $\delta_{NN}$, shift the energy level of the Hoyle resonance \cite{Ekstrom:2009ef},
\be
E_R = (0.2876 - 20.412 \delta_{NN} )~{\rm MeV},  
\ee
where the resonant energy is given with respect to the $^8$Be $+ \alpha$ threshold of 7.367 MeV.
In standard stellar evolutionary models for massive stars, 
most $^{12}$C is produced during the He burning phase. 
When the temperature becomes high enough, the $^{12}$C$(\alpha, \gamma){}^{16}$O reaction begins and $^{12}$C is processed to $^{16}$O. Massive stars end their He burning phases with a mixture of C and O. When $\delta_{NN} >0$, as would be expected for $\theta \ne 0$, $E_R$ is reduced, and the production of Carbon
becomes more efficient at a lower temperature. The burning of Carbon
to Oxygen does not occur and stars end their Helium burning phases with a core of almost pure Carbon.

If Oxygen is not present after He burning, there is little chance to 
subsequently produce it. Though some Oxygen is produced during 
Carbon burning, the Oxygen abundance in this phase of stellar evolution is reduced as Oxygen is processed to Ne through $\alpha$ capture. 
The analysis of Ref.~\cite{Ekstrom:2009ef} was 
based on stellar evolution models \cite{Ekstrom:2008gh} of 15 and 60 M$_\odot$, zero metallicity stars and found that for $\delta_{NN} \ge 0.3\,\%$, negligible amounts of Oxygen survive the 
Helium burning phase. 
Thus an upper limit of $\delta_{NN} < 0.002$ was set which corresponds to $B_d < 2.25$ MeV. 
This is a rather tight bound and corresponds to upper limits on  $\theta$ of 0.11 (0.11) for broken isospin, and 0.11 (0.12) for unbroken 
isospin for parameter Sets I (II), respectively. 
As shown above, the dineutron and the diproton remain unbound for such values of $\theta$, so that a universe with $0<\theta\lesssim 0.1$ will most probably look (almost) the same as a universe with $\theta=0$.

\section{Summary and conclusions}
\def\theequation{\arabic{section}.\arabic{equation}}
\label{sec:summary}

Let us summarize the pertinent results of our investigation for 
{ $0<\theta<\pi$}:
\begin{itemize}
\item As $\theta$ is increased, the deuteron is more strongly bound than in our world. 
  This means that for $\theta$ of the order one, there is much less fine-tuning than for $\theta=0$. Also, in the case of isospin symmetry, the values for the binding energy as $\theta$ approaches $\pi$ are compatible with calculations for the
  chiral limit. 
\item 
 The dineutron as well as the diproton are bound for $\theta\gtrsim 0.2$ and $\theta\gtrsim 0.7$, respectively. A bound diproton has often
   been considered a disaster for  the nucleosynthesis as we know it~\cite{Dyson}, but recent stellar calculations
   show that this might not be the case, see Ref.~\cite{MacDonald:2009vk,bradford,Barnes:2015mac}.
 \item Using Wigner's SU(4) symmetry and earlier results on systems with large scattering length, we have estimated
   the SU(4)-averaged binding energies of the three- and four-nucleon systems and found that these increase
   with increasing $\theta$ or with the deuteron binding energy. 
\item In general, we have found that nuclear binding energies are quite significantly altered when {$\theta=\order{1}$}.
While BBN would proceed, perhaps producing far less Helium and more Deuterium, changes in the deuteron binding energy would not
prevent the formation of stars and eventually life. Even a stable
diproton can not be excluded on this basis as stars would
continue to burn Hydrogen at lower temperatures. 
On the other hand, changes in the binding energy of $^8$Be and the resonant energy of the Hoyle state, would affect the triple $\alpha$ reaction rate
and lead to a world lacking in $^{16}$O.  
\item Applying the even stronger constraint not to upset the world as we enjoy it, we derived that $\theta$ must be $\lesssim 0.1$ in order to approximately recover the real nuclear reaction rates. In this case, the deviation of the neutron-proton mass difference to the real world value is less than 1\,\% and both the diproton and the dineutron still fail to be bound.
\end{itemize}

\section*{Acknowledgments}

MS is grateful to  Alexey Cherman and Maxim Pospelov for useful discussions.
UGM thanks Karlheinz Langanke, Maxim Mai and Andreas Wirzba for useful discussions.
This work was supported by DFG and NSFC through funds provided to the
Sino-German CRC 110 ``Symmetries and the Emergence of Structure in QCD" (NSFC
Grant No.~11621131001, DFG Grant No.~TRR110).
The work of UGM was supported in part by VolkswagenStiftung (Grant no. 93562)
and by the CAS President's International
Fellowship Initiative (PIFI) (Grant No.~2018DM0034).
The work of KO and MS is supported in part by U.S. Department of Energy (Grant No. DE-SC0011842)
and the work of DL is supported in part by the U.S. Department of Energy (Grant No. DE-SC0018638) 
and the NUclear Computational Low-Energy Initiative (NUCLEI) SciDAC project.

\appendix
\section{{\boldmath$\theta$}-dependence of the neutron-proton mass difference}
\def\theequation{\Alph{section}.\arabic{equation}}
\setcounter{equation}{0}
\label{sec:appA}

The strong contribution to the proton-neutron mass difference can be derived from the NLO $\pi N$ Lagrangian
\cite{Bernard:1995dp}
\begin{equation}\label{eq:massdiffLag}
\mathcal{L}^{\Delta m_N}_{\pi N} = \bar{N} c_5 \left(\chi_{+} - \frac{1}{2} \langle\chi_{+}\rangle\right) N~,
\end{equation}
where $c_5$ is a LEC, $N=(p,n)^{T}$ contains the nucleon fields, $\langle \ldots \rangle$ denotes the
trace in flavor space, and 
\begin{equation}
\chi_{+}= u^\dagger \chi_\theta u^\dagger + u \chi_\theta^\dagger u~.
\end{equation}
For the determination of the mass difference, $U=u^2$, which contains the pseudo-Nambu-Goldstone bosons of SU(2)
chiral perturbation theory, only needs to be expanded up to its leading order constant term. In particular, in a
$\theta$-vacuum $U$ is given by the vacuum alignment $U_0$. For $\chi_\theta=2 B \mathcal{M} \exp\left(i \theta/2\right)$,
with $\mathcal{M}=\operatorname{diag} \{m_u,m_d\}$ the quark mass matrix, we use the following parameterization
of the vacuum alignment:
\begin{equation}
U_0 = \operatorname{diag}\left\{e^{i \varphi}, e^{-i \varphi}\right\}~.
\end{equation}
Minimizing the vacuum energy density in SU(2) chiral perturbation theory (or equivalently removing the tree-level
tadpole term of the neutral pion), one finds \cite{Brower:2003yx}
\begin{equation}
\tan \varphi = -\varepsilon \tan \frac{\theta}{2}~,
\end{equation}
or
\begin{align}
  \sin \varphi & = \frac{-\varepsilon \tan\frac{\theta}{2}}{\sqrt{1+\varepsilon^2\tan^2\frac{\theta}{2}}}
  = \frac{-\varepsilon M_\pi^2 \sin\frac{\theta}{2}}{M_\pi^2(\theta)}~, \\
  \cos \varphi & = \frac{1}{\sqrt{1+\varepsilon^2\tan^2\frac{\theta}{2}}}
  = \frac{M_\pi^2 \cos\frac{\theta}{2}}{M_\pi^2(\theta)}~,
\end{align}
where we have used Eq.~\eqref{eq:mpithetaiso}. With that, Eq.~\eqref{eq:massdiffLag} becomes
\begin{align}
\mathcal{L}^{\Delta m_N}_{\pi N} & =  \bar{N} 4 c_5 B_0 \frac{m_u \cos\left(\frac{\theta}{2}-\varphi\right)-m_d\cos\left(\frac{\theta}{2}+\varphi\right)}{2} \tau_3 N \nonumber\\
& = \bar{N} 4 c_5 B_0 \frac{M_\pi^2}{M_\pi^2(\theta)} \frac{m_u-m_d}{2} \tau_3 N~,
\end{align}
which results in the strong contribution to the proton-neutron mass difference given in Eq.~\eqref{eq:deltamnp}.


\end{document}